\pdfoutput=1
\documentclass[12pt,a4paper]{article}

\usepackage{ifthen} 
\newboolean{pdflatex}
\setboolean{pdflatex}{true} 

\newboolean{articletitles}
\setboolean{articletitles}{true} 

\newboolean{uprightparticles}
\setboolean{uprightparticles}{false} 

\def\paperauthors{LHCb collaboration}
\def\paperasciititle{Measurement of the charm mixing parameter yCP-yCPKpi in prompt D0 meson decays} 
\def\papertitle{Measurement of the charm mixing parameter $y_{C\!P} - y_{C\!P}^{K\pi}$ using two-body $D^0$ meson decays } 
\def\paperkeywords{{High Energy Physics}, {LHCb}, {flavour physics}, {charm physics}, {CP violation}, {time-dependent CP violation}, {indirect CP violation}, {oscillation}}
\def\papercopyright{\the\year\ CERN for the benefit of the LHCb collaboration}
\def\paperlicence{CC BY 4.0 licence}
\def\paperlicenceurl{https://creativecommons.org/licenses/by/4.0/}

\usepackage[top=1in, bottom=1.25in, left=1in, right=1in]{geometry}

\usepackage{microtype}
\usepackage{lineno}  
\usepackage{xspace} 
\usepackage{caption} 

\usepackage{graphicx}  
\usepackage{color}
\usepackage{colortbl}
\graphicspath{{./figs/}{./figs_supplementary/}} 

\usepackage{amsmath} 
\usepackage{amssymb}
\usepackage{amsfonts}
\usepackage{upgreek} 

\newcommand*\patchAmsMathEnvironmentForLineno[1]{%
\expandafter\let\csname old#1\expandafter\endcsname\csname #1\endcsname
\expandafter\let\csname oldend#1\expandafter\endcsname\csname
end#1\endcsname
 \renewenvironment{#1}%
   {\linenomath\csname old#1\endcsname}%
   {\csname oldend#1\endcsname\endlinenomath}%
}
\newcommand*\patchBothAmsMathEnvironmentsForLineno[1]{%
  \patchAmsMathEnvironmentForLineno{#1}%
  \patchAmsMathEnvironmentForLineno{#1*}%
}
\AtBeginDocument{%
\patchBothAmsMathEnvironmentsForLineno{equation}%
\patchBothAmsMathEnvironmentsForLineno{align}%
\patchBothAmsMathEnvironmentsForLineno{flalign}%
\patchBothAmsMathEnvironmentsForLineno{alignat}%
\patchBothAmsMathEnvironmentsForLineno{gather}%
\patchBothAmsMathEnvironmentsForLineno{multline}%
\patchBothAmsMathEnvironmentsForLineno{eqnarray}%
}

\usepackage{hyperxmp}

\usepackage[pdftex,
            pdfauthor={\paperauthors},
            pdftitle={\paperasciititle},
            pdfkeywords={\paperkeywords},
            pdfcopyright={Copyright (C) \papercopyright},
            pdflicenseurl={\paperlicenceurl}]{hyperref}

\usepackage[colorinlistoftodos,textsize=scriptsize]{todonotes}

\usepackage[bottom,flushmargin,hang,multiple]{footmisc}

\usepackage[all]{hypcap} 

\usepackage{xspace} 
\usepackage{upgreek}

\def\lhcb   {\mbox{LHCb}\xspace}

\def\rich   {RICH\xspace}

\def\MagUp {\mbox{\em Mag\kern -0.05em Up}\xspace}
\def\MagDown {\mbox{\em MagDown}\xspace}

\ifthenelse{\boolean{uprightparticles}}%
{

 \def\Ppi         {\ensuremath{\uppi}\xspace}

 \def\PDelta      {\ensuremath{\Delta}\xspace}                 
 \def\PXi         {\ensuremath{\Xi}\xspace}                 
 \def\PLambda     {\ensuremath{\Lambda}\xspace}                 
 \def\PSigma      {\ensuremath{\Sigma}\xspace}                 
 \def\POmega      {\ensuremath{\Omega}\xspace}                 
 \def\PUpsilon    {\ensuremath{\Upsilon}\xspace}

 \def\PB      {\ensuremath{\mathrm{B}}\xspace}                 
                  
 \def\PD      {\ensuremath{\mathrm{D}}\xspace}

 \def\PK      {\ensuremath{\mathrm{K}}\xspace}

 \def\Pb      {\ensuremath{\mathrm{b}}\xspace}                 
 \def\Pc      {\ensuremath{\mathrm{c}}\xspace}

 \def\Ph      {\ensuremath{\mathrm{h}}\xspace}                 
 \def\Pi      {\ensuremath{\mathrm{i}}\xspace}

 \def\Pp      {\ensuremath{\mathrm{p}}\xspace}

 \def\Ps      {\ensuremath{\mathrm{s}}\xspace}

 \def\thebaroffset{0.0em}
}
{

 \def\Ppi         {\ensuremath{\pi}\xspace}

 \mathchardef\PDelta="7101
 \mathchardef\PXi="7104
 \mathchardef\PLambda="7103
 \mathchardef\PSigma="7106
 \mathchardef\POmega="710A
 \mathchardef\PUpsilon="7107
                  
 \def\PB      {\ensuremath{B}\xspace}                 
                  
 \def\PD      {\ensuremath{D}\xspace}

 \def\PK      {\ensuremath{K}\xspace}

 \def\Pb      {\ensuremath{b}\xspace}                 
 \def\Pc      {\ensuremath{c}\xspace}

 \def\Ph      {\ensuremath{h}\xspace}                 
 \def\Pi      {\ensuremath{i}\xspace}

 \def\Pp      {\ensuremath{p}\xspace}

 \def\Ps      {\ensuremath{s}\xspace}

 \def\thebaroffset{0.18em}
}
\newcommand{\offsetoverline}[2][\thebaroffset]{\kern #1\overline{\kern -#1 #2}}%

\makeatletter
\ifcase \@ptsize \relax
  \newcommand{\miniscule}{\@setfontsize\miniscule{4}{5}}
\or
  \newcommand{\miniscule}{\@setfontsize\miniscule{5}{6}}
\or
  \newcommand{\miniscule}{\@setfontsize\miniscule{5}{6}}
\fi
\makeatother

\DeclareRobustCommand{\optbar}[1]{\shortstack{{\miniscule (\rule[.5ex]{1.25em}{.18mm})}
  \\ [-.7ex] $#1$}}

\def\squark    {{\ensuremath{\Ps}}\xspace}

\def\cquark    {{\ensuremath{\Pc}}\xspace}

\def\bquark    {{\ensuremath{\Pb}}\xspace}

\def\hadron {{\ensuremath{\Ph}}\xspace}
\def\pion   {{\ensuremath{\Ppi}}\xspace}

\def\pip    {{\ensuremath{\pion^+}}\xspace}
\def\pim    {{\ensuremath{\pion^-}}\xspace}

\def\kaon    {{\ensuremath{\PK}}\xspace}

\def\KorKbar {\kern \thebaroffset\optbar{\kern -\thebaroffset \PK}{}\xspace}

\def\Kp      {{\ensuremath{\kaon^+}}\xspace}
\def\Km      {{\ensuremath{\kaon^-}}\xspace}

\def\Dbar    {{\ensuremath{\offsetoverline{\PD}}}\xspace}
\def\D       {{\ensuremath{\PD}}\xspace}

\def\DorDbar {\kern \thebaroffset\optbar{\kern -\thebaroffset \PD}\xspace}
\def\Dz      {{\ensuremath{\D^0}}\xspace}
\def\Dzb     {{\ensuremath{\Dbar{}^0}}\xspace}
\def\Dp      {{\ensuremath{\D^+}}\xspace}
\def\Dm      {{\ensuremath{\D^-}}\xspace}

\def\DpDm    {\ensuremath{\Dp {\kern -0.16em \Dm}}\xspace}

\def\Dstarp  {{\ensuremath{\D^{*+}}}\xspace}

\def\theDstarp{{\ensuremath{\D^{*}(2010)^{+}}}\xspace}

\def\B       {{\ensuremath{\PB}}\xspace}

\def\BorBbar {\kern \thebaroffset\optbar{\kern -\thebaroffset \PB}\xspace}

\def\Bd      {{\ensuremath{\B^0}}\xspace}

\def\BdorBdbar {\kern \thebaroffset\optbar{\kern -\thebaroffset \Bd}\xspace}

\def\Bs      {{\ensuremath{\B^0_\squark}}\xspace}

\def\BsorBsbar {\kern \thebaroffset\optbar{\kern -\thebaroffset \Bs}\xspace}

\def\Y#1S{\ensuremath{\PUpsilon{(#1S)}}\xspace}

\def\proton      {{\ensuremath{\Pp}}\xspace}

\def\LorLbar     {\kern \thebaroffset\optbar{\kern -\thebaroffset \PLambda}\xspace}

\newcommand{\decay}[2]{\ensuremath{#1\!\to #2}\xspace} 
\def\ra                 {\ensuremath{\rightarrow}\xspace}
\def\to                 {\ensuremath{\rightarrow}\xspace}

\def\CP                {{\ensuremath{C\!P}}\xspace}

\def\AT#1     {\ensuremath{A_{\mathrm{T}}^{#1}}\xspace}           

\def\C#1      {\ensuremath{\mathcal{C}_{#1}}\xspace}                       
\def\Cp#1     {\ensuremath{\mathcal{C}_{#1}^{'}}\xspace}                    
\def\Ceff#1   {\ensuremath{\mathcal{C}_{#1}^{\mathrm{(eff)}}}\xspace}        
\def\Cpeff#1  {\ensuremath{\mathcal{C}_{#1}^{'\mathrm{(eff)}}}\xspace}       
\def\Ope#1    {\ensuremath{\mathcal{O}_{#1}}\xspace}                       
\def\Opep#1   {\ensuremath{\mathcal{O}_{#1}^{'}}\xspace}                    

\def\ycp        {\ensuremath{y_{\CP}}\xspace}

\newcommand{\nospaceunit}[1]{\ensuremath{\text{#1}}}       
\newcommand{\aunit}[1]{\ensuremath{\text{\,#1}}}       

\newcommand{\tev}{\aunit{Te\kern -0.1em V}\xspace}
\newcommand{\gev}{\aunit{Ge\kern -0.1em V}\xspace}
\newcommand{\mev}{\aunit{Me\kern -0.1em V}\xspace}
\newcommand{\kev}{\aunit{ke\kern -0.1em V}\xspace}
\newcommand{\ev}{\aunit{e\kern -0.1em V}\xspace}
 
\newcommand{\mevc}{\ensuremath{\aunit{Me\kern -0.1em V\!/}c}\xspace}
\newcommand{\gevc}{\ensuremath{\aunit{Ge\kern -0.1em V\!/}c}\xspace}
\newcommand{\mevcc}{\ensuremath{\aunit{Me\kern -0.1em V\!/}c^2}\xspace}
\newcommand{\gevcc}{\ensuremath{\aunit{Ge\kern -0.1em V\!/}c^2}\xspace}

\def\cm   {\aunit{cm}\xspace}

\def\mm   {\aunit{mm}\xspace}

\def\mum  {\ensuremath{\,\upmu\nospaceunit{m}}\xspace}

\def\fb   {\ensuremath{\aunit{fb}}\xspace}
\def\invfb   {\ensuremath{\fb^{-1}}\xspace}

\def\fs   {\aunit{fs}}

\def\gsim{{~\raise.15em\hbox{$>$}\kern-.85em
          \lower.35em\hbox{$\sim$}~}\xspace}
\def\lsim{{~\raise.15em\hbox{$<$}\kern-.85em
          \lower.35em\hbox{$\sim$}~}\xspace}

\def\pt         {\ensuremath{p_{\mathrm{T}}}\xspace}

\def\ptot       {\ensuremath{p}\xspace}

\def\mrad{\aunit{mrad}\xspace}

\def\evtgen     {\mbox{\textsc{EvtGen}}\xspace}

\def\geant      {\mbox{\textsc{Geant4}}\xspace}

\def\photos     {\mbox{\textsc{Photos}}\xspace}

\def\pythia     {\mbox{\textsc{Pythia}}\xspace}
\def\rapidsim    {\mbox{\textsc{RapidSim}}\xspace}

\def\tell1  {TELL1\xspace}
\def\ukl1   {UKL1\xspace}

\input{my-symbols-def}

\usepackage{cite} 
\usepackage{mciteplus}

\begin{document}

\renewcommand{\thefootnote}{\fnsymbol{footnote}}
\setcounter{footnote}{1}


\begin{titlepage}
\pagenumbering{roman}

\vspace*{-1.5cm}
\centerline{\large EUROPEAN ORGANIZATION FOR NUCLEAR RESEARCH (CERN)}
\vspace*{1.5cm}
\noindent
\begin{tabular*}{\linewidth}{lc@{\extracolsep{\fill}}r@{\extracolsep{0pt}}}
\ifthenelse{\boolean{pdflatex}}
{\vspace*{-1.5cm}\mbox{\!\!\!\includegraphics[width=.14\textwidth]{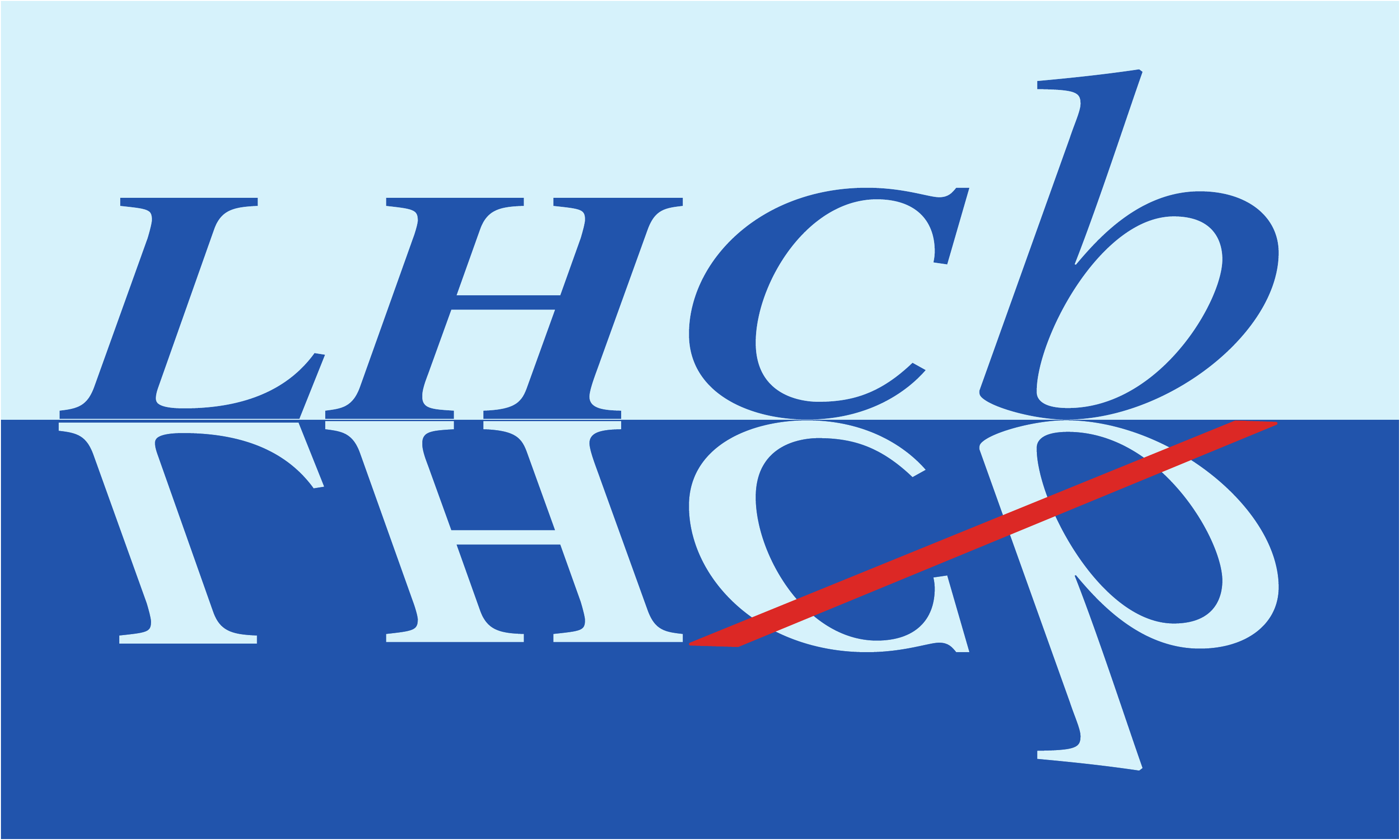}} & &}
{\vspace*{-1.2cm}\mbox{\!\!\!\includegraphics[width=.12\textwidth]{lhcb-logo.eps}} & &}
\\
 & & CERN-EP-2022-022 \\
 & & LHCb-PAPER-2021-041 \\
 & & May 30, 2022 \\
\end{tabular*}

\vspace*{2.5cm}

{\normalfont\bfseries\boldmath\huge
\begin{center}
  \papertitle 
\end{center}
}

\vspace*{1.0cm}

\begin{center}
\paperauthors\footnote{Authors are listed at the end of this paper.}
\end{center}

\vspace{\fill}

\begin{abstract}
  \noindent
    A measurement of the ratios of the effective decay widths of \DzPP and \DzKK decays over that of \DzRS decays is performed with the \lhcb experiment using proton--proton collisions at a centre-of-mass energy of $13\tev$, corresponding to an integrated luminosity of $6\invfb$. 
    These observables give access to the charm mixing parameters $\ycp^{\pi\pi} - \ycp^{K\pi}$ and $\ycp^{KK} - \ycp^{K\pi}$, and are measured as
    \begin{equation*}
    \begin{split}
        \ycp^{\pi\pi} - \ycp^{K\pi} & = (6.57 \pm 0.53 \pm 0.16) \times 10^{-3} \, , \\
    \ycp^{KK} - \ycp^{K\pi} & = (7.08 \pm 0.30 \pm 0.14) \times 10^{-3} \, ,
    \end{split}
    \end{equation*}
    where the first uncertainties are statistical and the second systematic. The combination of the two measurements is  \mbox{$\ycp - \ycp^{K\pi} = (6.96 \pm 0.26 \pm 0.13) \times 10^{-3}$}, which is four times more precise than the previous world average. 
\end{abstract}

\vspace*{2.0cm}

\begin{center}
  Published in \textit{Phys. Rev.} \textbf{D105} (2022) 092013
\end{center}

\vspace{\fill}

{\footnotesize 
\centerline{\copyright~\papercopyright. \href{\paperlicenceurl}{\paperlicence}.}}
\vspace*{2mm}

\end{titlepage}


\newpage
\setcounter{page}{2}
\mbox{~}

\renewcommand{\thefootnote}{\arabic{footnote}}
\setcounter{footnote}{0}

\cleardoublepage


\pagestyle{plain} 
\setcounter{page}{1}
\pagenumbering{arabic}


\section{Introduction}
\label{sect:introduction}

Neutral charm mesons can change their flavour and turn into their antimeson counterpart before they decay. This phenomenon, known as \Dz--\Dzb mixing, does not occur at tree level in the Standard Model and is sensitive to contributions from new particles arising in extensions of the Standard Model. 
The mass eigenstates of neutral charm mesons can be expressed as a linear combination of their flavour eigenstates, $|D_{1,2}\rangle = p |\Dz\rangle \pm q |\Dzb\rangle$, where $p$ and $q$ are complex parameters satisfying $|p|^2 + |q|^2 = 1$. In the limit of charge-parity (\CP) symmetry, the relation $|q/p|=1$ holds.
The time evolution of neutral charm meson systems is governed by the effective Hamiltonian $\mathbf{H} = \mathbf{M} - \frac{i}{2}\mathbf{\Gamma}$, where the Hermitian matrices $\mathbf{M}$ and $\mathbf{\Gamma}$ describe $(\Dz,\Dzb) \leftrightarrow (\Dz, \Dzb)$ dispersive transitions through virtual intermediate states and absorptive transitions through real intermediate states, respectively~\cite{PDG2020}. 
The \Dz--\Dzb oscillations are described by the two dimensionless parameters $x_{12}=2|M_{12}/\Gamma|$ and $y_{12}=|\Gamma_{12}/\Gamma|$~\cite{Grossman:2009mn,Kagan:2009gb}, where $\Gamma = (\Gamma_1 + \Gamma_2)/2$ is the average decay width of the $D_1$ and $D_2$ states, and $M_{12}$ ($\Gamma_{12}$) is the off-diagonal element of matrix $\mathbf{M}$ ($\mathbf{\Gamma}$). The values of $x_{12}$ and $y_{12}$ are of the order of half a percent and have been measured to be significantly different from zero~\cite{delAmoSanchez:2010xz,Lees:2012qh,Peng:2014oda,Staric:2015sta,LHCb-PAPER-2017-046,LHCb-PAPER-2018-038,LHCb-PAPER-2021-009,LHCb-PAPER-2021-033}.

The non-zero value of~$y_{12}$ implies that the time-dependent decay rate of Cabibbo-suppressed \DzF decays, with $f=K^-K^+,\pi^-\pi^+$ final states, is described by an exponential function with an effective decay width $\hat{\Gamma}$ that differs slightly from $\Gamma$.
The departure from unity of the ratio of the effective decay widths of \DzPP and \DzKK decays over that of \DzRS decays is measured via the observable~\cite{PDG2020}
\begin{equation}
   \ycp^{f}  = \frac{\hat{\Gamma}(\DzF) + \hat{\Gamma}(\DzbF)}{2 \Gamma} - 1  \, .
\label{eq:yCP_CS}
\end{equation}
The above quantity can be approximated as~\cite{Kagan:2020vri} 
\begin{equation}
    \ycp^{f} = y_{12}\cos \phi_f^{\Gamma} \, ,
\end{equation}
where $\phi_f^{\Gamma} = \mathrm{arg}\left(\Gamma_{12}A_f/\overline{A}_f \right)$ describes the  \CP-violating phase difference of the interference between decay amplitudes with and without absorptive mixing~\cite{Grossman:2009mn,Kagan:2009gb}, and $A_f$ ($\overline{A}_f$) is the decay amplitude of a $\Dz$ ($\Dzb$) meson to the final state~$f$. Any deviation of $\ycp^f$ from $y_{12}$ would be a sign of \CP violation. At the current experimental sensitivity, final-state dependent contributions to $\ycp^f$ can be neglected in the limit where the phase $\phi_f^{\Gamma}$ is replaced by the universal phase $\phi_2^{\Gamma}$, and \mbox{$\ycp \approx y_{12}\cos\phi_2^{\Gamma}$}~\cite{Kagan:2020vri}.  The parameter $y_{12}$ is equal to $|y| \equiv |\Gamma_1 - \Gamma_2|/2\Gamma$ up to second order \CP violation effects~\cite{Kagan:2020vri}, where the best experimental estimate is $y = (6.30^{+0.33}_{-0.30})\times 10^{-3}$~\cite{LHCb-PAPER-2021-033}. The current world average gives $\phi_2^\Gamma = (48^{+29}_{-28}) \mrad$~\cite{Pajero_Morello_paper,charm-fitter-tommaso}, implying that $|y_{12} - \ycp| < 3 \times 10^{-5}$ at $95\%$ confidence level. Since this upper limit is about one order of magnitude smaller than the current experimental sensitivity on both $y_{12}$ and \ycp at \lhcb, an accurate measurement of \ycp provides important constraints on $y_{12}$.

The previous measurements of \ycp performed by the BABAR~\cite{Lees:2012qh}, Belle~\cite{Staric:2015sta} and \mbox{\lhcb~\cite{LHCb-PAPER-2011-032,LHCb-PAPER-2018-038}} collaborations use the average decay width of \DzRS and $\Dzb \to K^+\pi^-$ decays as a proxy to the decay width $\Gamma$. It was recently shown in Ref.~\cite{Pajero_Morello_paper} that the use of this proxy inside the experimental observable of Eq.~\eqref{eq:yCP_CS} does not give direct access to $\ycp^f$ but rather corresponds to
\begin{equation}
  \frac{\hat{\Gamma}(\DzF) + \hat{\Gamma}(\DzbF)}{\hat{\Gamma}(\DzRS) + \hat{\Gamma}(\DzbRS)} - 1 \approx \ycp^f - \ycp^{K\pi} \, . 
\end{equation}
The quantity $\ycp^{K\pi}$ is approximately equal to
\begin{equation}
    \ycp^{K\pi} \approx \sqrt{R_D}\left( x_{12}\cos\phi_2^M\sin\delta_{K\pi} + y_{12} \cos\phi_2^\Gamma \cos \delta_{K\pi} \right) \approx -0.4 \times 10^{-3}\, ,
\end{equation}
where $R_D$ is the ratio of the branching fractions of the doubly Cabibbo-suppressed \DzWS decay over the Cabibbo-favoured \DzRS decay. The current best experimental estimate is $\sqrt{R_D} = (5.87 \pm 0.02)\times 10^{-2}$~\cite{HFLAV18}. The phase $\phi^{M}_{2}$ is equal to the phase of $M_{12}$ with respect to its $\Delta U = 2$ dominant contribution, and $\delta_{K\pi}$ is the strong-phase difference between the doubly Cabibbo-suppressed and Cabibbo-favoured decay amplitudes~\cite{LHCb-PAPER-2021-033}. In the limit of no \CP violation and of $U$-spin symmetry in $\Dz \to K^{\mp}\pi^{\pm}$ decays, the approximations 
$\delta_{K\pi} \approx \pi$
and $\ycp - \ycp^{K\pi} \approx y_{12}(1 + \sqrt{R_D})$ hold.

The world average value of $\ycp - \ycp^{K\pi}$ is measured to be  $(7.19 \pm 1.13) \times 10^{-3}$~\cite{HFLAV18}. This paper reports a new measurement of  $\ycp - \ycp^{K\pi}$. The result is obtained from a weighted average of statistically independent measurements with $K^-K^+$ and $\pi^-\pi^+$ final states, using proton-proton ($pp$) collision data collected with the \lhcb experiment at a centre-of-mass energy of $13\tev$ in the Run~2 data taking period (2015--2018), corresponding to an integrated luminosity of $6\invfb$. The $\Dz$ mesons are required to originate from $D^{\ast}(2010)^+ \ra \Dz \pi^+_{\mathrm{tag}}$ decays, such that their flavour at production is identified by the charge of the tagging pion, $\pi^+_{\mathrm{tag}}$. The inclusion of charge-conjugate processes is implied throughout. Hereafter the $D^{\ast}(2010)^+$ meson is referred to as a $D^{\ast+}$ meson. 
\section{\lhcb detector }
\label{sect:detector}

The \lhcb detector~\cite{LHCb-DP-2008-001,LHCb-DP-2014-002} is a single-arm forward spectrometer covering the pseudorapidity range $2<\eta <5$, designed for the study of particles containing \bquark or \cquark quarks.
The detector includes a high-precision tracking system consisting of a silicon-strip vertex detector surrounding the $pp$ interaction region, a large-area silicon-strip detector located upstream of a dipole magnet with a bending power of about $4{\mathrm{\,Tm}}$, and three stations of silicon-strip detectors and straw drift tubes placed downstream of the magnet.
The tracking system provides a measurement of the momentum, \ptot, of charged particles with a relative uncertainty varying from 0.5\% at low momentum to 1.0\% at 200\gevc.
The minimum distance of a track to a primary vertex (PV), the impact parameter (IP), is measured with a resolution of $(15+29/\pt)\mum$, where \pt is the component of the momentum transverse to the beam, in\,\gevc.
The \lhcb coordinate system is right-handed, with the~$z$ axis pointing along the beam axis, $y$ the vertical direction pointing upwards, and $x$ the horizontal direction.
The origin corresponds to the nominal \proton\proton interaction point. 
The magnetic field deflects oppositely charged particles in opposite directions along the $x$ axis, inducing potential detection asymmetries.
Therefore, the magnet polarity is reversed regularly throughout the data taking to reduce the effects of detection asymmetries. The two polarities are referred to as \textit{MagUp} and \textit{MagDown}.
Different types of charged hadrons are distinguished using information from two ring-imaging Cherenkov (\rich) detectors.
Photons, electrons and hadrons are identified by a calorimeter system consisting of scintillating-pad and preshower detectors, an electromagnetic and a hadronic calorimeter. Muons are identified by a system composed of alternating layers of iron and multiwire proportional chambers.

The online event selection is performed by a trigger, which consists of a hardware stage followed by a two-level software stage, which applies a full event reconstruction.
The good performance of the online reconstruction allows this measurement to be performed using candidates reconstructed directly at the trigger level~\cite{LHCb-DP-2012-004,LHCb-DP-2016-001}.

Simulation is used to study the background of secondary $D^{\ast+}$ candidates from $B$ meson decays (Sect.~\ref{sect:backgrounds}), and to validate the analysis procedure. The \proton\proton collisions are generated with \pythia~\cite{Sjostrand:2007gs} with a specific \lhcb configuration~\cite{LHCb-PROC-2010-056}. The interaction of the simulated particles with the detector material are described using the \geant toolkit~\cite{Allison:2006ve,LHCb-PROC-2011-006}. Decays of unstable particles are described by \evtgen~\cite{evtgen}, in which final state radiation is generated using \photos~\cite{PHOTOS}. In addition, fast simulation is generated with the \rapidsim package~\cite{Cowan:2016tnm}. \rapidsim simulations allow for a first validation of the analysis procedure (Sect.~\ref{sect:validation_simulation}), and for a description of the background under the $\Dz$ mass peak (Sect.~\ref{sect:systematics}).

\section{Measurement strategy}
\label{sect:analysis_strategy}

The parameters $\ycp^{f} - \ycp^{K\pi}$ are measured from the decay-time ratios $R^f(t)$ of \DzF over \DzRS signal yields as a function of the reconstructed $\Dz$ decay time, $t$, assuming all $\Dz$ mesons are produced at the PV,
\begin{equation}
    R^{f}(t) = \frac{N(\DzF,t)}{N(\DzRS,t)} \propto e^{-(\ycp^f - \ycp^{K\pi}) \, t/\tau_{\Dz}} \frac{\varepsilon(f,t)}{\varepsilon(K^-\pi^+,t)} \, , 
\label{eq:Rraw_t}
\end{equation}
where $\tau_{\Dz} = (410.1 \pm 1.5) \fs$ is the measured lifetime of the $\Dz$ meson~\cite{PDG2020}, and $\varepsilon(h^-h^{'+},t)$, with $h^{(')\pm}$ denoting $K^{\pm}$ or $\pi^{\pm}$, is the time-dependent efficiency for the considered final state. Equation~\eqref{eq:Rraw_t} indicates that the access to $\ycp^f - \ycp^{K\pi}$ using an exponential fit is affected by the presence of both efficiencies. 
In this paper, the term \textit{numerator (denominator) decay} refers to the decay quoted in the numerator (denominator) of the ratio~$R^{f}(t)$.
The time-dependent efficiency can be written as the product of two distinct components.
The selection efficiency is related to requirements applied at various stages of the \lhcb data acquisition system, while the detection efficiency arises from the interaction of the charged kaons and pions with the \lhcb detector.
The time dependence of the efficiencies of the numerator and denominator decays differs because of their different final states, and could bias the measurement if not accounted for.
The analysis strategy consists of equalising the selection efficiencies and then the detection efficiencies of the numerator and denominator decays. Their combined effects cancel out in the decay time ratio, such that $\ycp^f - \ycp^{K\pi}$ can be measured  without additional corrections. Both steps are performed using data-driven methods detailed in the following paragraphs.

\begin{figure}[tb]
    \centering
    \includegraphics[width=\linewidth]{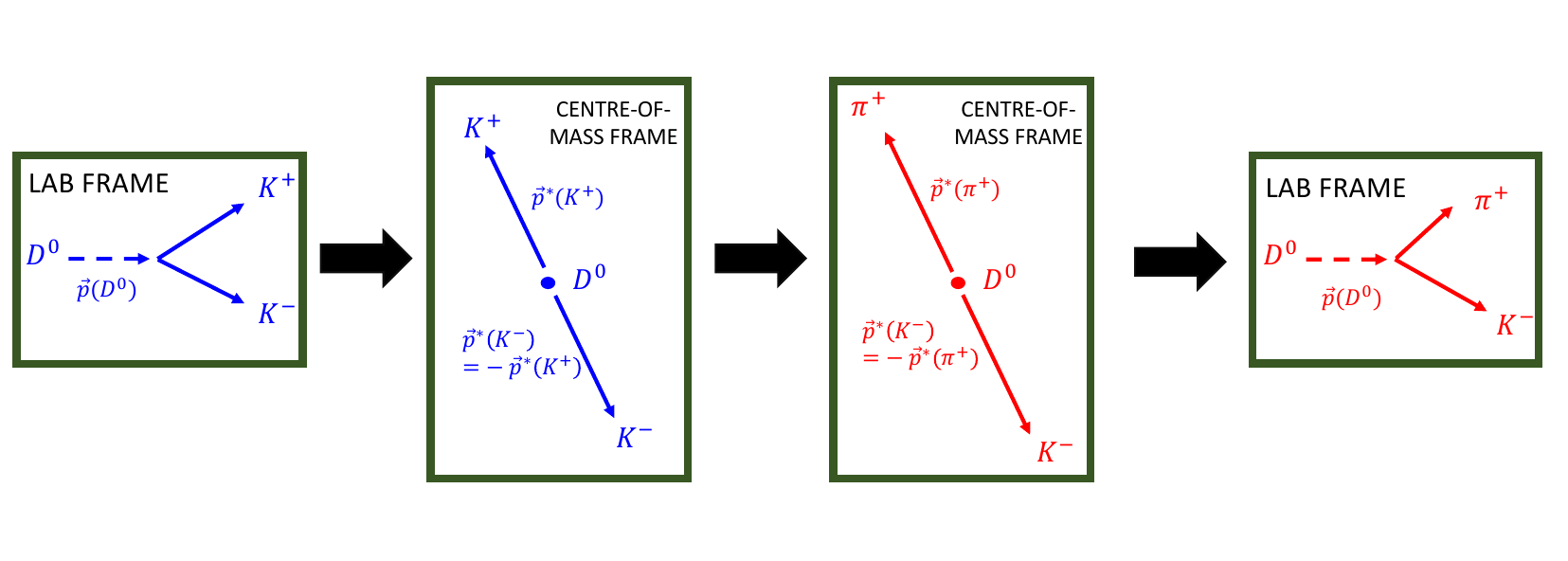}
    \caption{Sketch of a \DzKK to \DzRS matching.}
    \label{fig:Matching_Drawing}
\end{figure}

The selection efficiencies of \DzF and \DzRS decays mainly differ because of the different masses of their final-state particles, leading to distinct kinematic distributions of the final state particles of the $\Dz$ candidate in the laboratory (lab) frame.
The parent $\Dz$ meson has a momentum $p$ and decay angle $\theta^{\ast}(h^-)$
that are independent of the pair of the final states considered in this analysis.
The angle $\theta^{\ast}(h^-)$ is defined as the angle between the momentum of the negatively charged final state particle $h^-$ in the centre-of-mass frame of the $\Dz$ meson and the $\Dz$ momentum in the lab frame.
To obtain equal acceptance for both decays, we require that each $\Dz$ candidate selected in one final state would also pass the selection requirements for the other final state with the same $\Dz$ kinematic properties.
A \textit{kinematic matching procedure} has been developed for this purpose~\cite{Pietrzyk:2803301}. It consists of an event-by-event analytical transformation, which matches the final-state kinematic variables of one decay to the other. To match the kinematics of a \DzKK decay to a \DzRS decay (sketched in Fig.~\ref{fig:Matching_Drawing}),
a boost to the centre-of-mass frame of the $\Dz$ candidate is performed, such that both final-state particle momenta have equal magnitude,
\begin{equation}
    |\vec{p}^{\ast}| = \frac{\sqrt{(m_{\Dz}^2-(m_{K^+}-m_{K^-})^2)(m_{\Dz}^2-(m_{K^+}+m_{K^-})^2)}}{2m_{\Dz}}\,,
\end{equation}
where $m_i$ refers to the masses of the particles. 
By substituting $m_{K^+}$ with $m_{\pi^+}$, $|p^{\ast}|$  changes from $791\mevc$ to $861\mevc$, and a \DzRS state with identical kinematic properties is generated. The use of the $K^-\pi^+$ kinematics in the lab frame derived from this procedure (referred to as \textit{matched} kinematic quantities) ensures that both the matched \DzKK and the target \DzRS decays cover the same kinematic phase space.

The correction of the difference of detection efficiencies is treated with the \textit{kinematic weighting procedure}, which is performed after the kinematic matching. The procedure consists of weighting the $p$, $\pt$ and $\eta$ distributions of the $D^{\ast+}$ meson and both matched final-state particles of one of the decays to the distributions of the other decay. The procedure is performed using a gradient-boosted-reweighting algorithm from the \texttt{hep\_ml} library~\cite{hep-ml}.

The analysis procedure is validated with three distinct methods. First, a measurement of $\ycp^{KK} - \ycp^{K\pi}$ is performed making use of fast simulation samples generated with the \rapidsim package, where  strong variations of the time-dependent efficiencies as a function of the kinematic variables are introduced to test the robustness of the procedure. 
Second, the measurement is performed making use of large fully simulated samples. 
Finally, the procedure is validated with \lhcb data through a study of a cross-check observable, $R^{CC}(t)$, built from the time-dependent ratio of the yields of \DzPP and \DzKK decays,
\begin{equation}
    R^{CC}(t) = \frac{N(\DzPP,t)}{N(\DzKK,t)} \propto e^{-\ycp^{CC} \, t/\tau_{\Dz}} \frac{\varepsilon(\pi^-\pi^+,t)}{\varepsilon(K^-K^+,t)} \, ,
\label{eq:RCCraw_t}
\end{equation}
where the parameter $\ycp^{CC}$ is expected to be compatible with zero, since the final-state dependent part of $\ycp$ is negligible.
The observable $R^{CC}(t)$ benefits from the fact that both final state tracks are different for numerator and denominator decays, increasing the biasing effects from their corresponding efficiencies. 

The data samples are contaminated by the presence of three noticeable background contributions. The first is the combinatorial background, which is subtracted by means of a fit to the distribution of $\Delta m = m(h^-h^+ \pi^+_{\mathrm{tag}})-m(h^-h^+)$, where $m(h^-h^+ \pi^+_{\mathrm{tag}})$ is the mass of the $D^{\ast +}$ candidate and $m(h^-h^+)$ that of the \Dz candidate.
The second background contribution comes from $D^{\ast+}$ mesons that are not produced at the PV but from the decay of $B$ mesons. The effect of such secondary decays on the measurement is accounted by including their presence in the fit model of Eq.~\eqref{eq:Rraw_t}. The treatment of the combinatorial background and of secondary decays is detailed in Sect.~\ref{sect:backgrounds}.
A third background contribution is related to the presence of partially reconstructed or misreconstructed $D^{\ast+} \ra \Dz \pi^+$ decays. A systematic uncertainty is estimated to cover their impact on the measurement and is discussed in Sect.~\ref{sect:systematics}. 

\section{Candidate selection}
\label{sect:selection}

The $D^{\ast +} \ra (D^0 \to h^-h^{'+}) \pi_{\mathrm{tag}}^+$ decays are reconstructed at the trigger level. 
At the hardware stage, the trigger decision is required to be based on particles independent of the signal candidates, as requiring a decision depending on the signal candidates would degrade the performance of the kinematic matching procedure.
Both software trigger stages were specifically designed to minimise the biasing effects to the decay time ratio $R^{f}(t)$, as detailed in Ref.\cite{LHCb-PUB-2015-026}. This is achieved by avoiding requirements on kinematic variables of the final-state particles that are strongly correlated with the $\Dz$ decay time.
Candidate $\Dz$ mesons are constructed from $h^-h^{'+}$ pairs which have a distance of closest approach of less than $100\mum$, form a vertex with a $\chi^2$ per degree of freedom less than ten, and have an invariant mass in the interval $[1804,1924]\mevcc$.
The reconstructed $\Dz$ decay time is required to be higher than $0.6 \, \tau_{\Dz}$.
The angle between the $\Dz$ momentum vector and the vector connecting the $\Dz$ decay vertex and the PV is required to be less than $8^{\circ}$, and the $\Dz$ transverse momentum larger than  $2\gevc$.
Both final-state particles are required to have an individual transverse momentum above $800\mevc$, and at least one of these must have a transverse momentum exceeding $1200\mevc$. Furthermore, their individual absolute momenta are required to be higher than $5\gevc$. 
Finally, based on the information provided by the RICH detectors, the final-state candidates are assigned a pion or kaon mass. 
To remove statistical correlations between the $\ycp^{\pi\pi} - \ycp^{K\pi}$ and $\ycp^{KK} - \ycp^{K\pi}$ measurements related to the common $K^-\pi^+$ final state, the \DzRS sample is split into two statistically independent samples. Since three times more \DzKK than \DzPP signal candidates are selected, the \DzRS sample is split accordingly for the  $\ycp^{\pi\pi} - \ycp^{K\pi}$ and $\ycp^{KK} - \ycp^{K\pi}$ measurements.

In the offline selection, all kaon and pion tracks are required to have a pseudorapidity in the range $2.0$ to $4.2$ to remove particles traversing regions of high material density. The $\Dz$ flight distance in the $x-y$ plane is required to be less than $4\mm$ to remove $D^{\ast+}$ candidates produced from interactions with the detector material. The $z$-coordinate of the $\Dz$ decay vertex is required not to exceed a distance of $20\cm$ from the \proton\proton interaction point to remove residual background reconstructed at larger distances. The invariant mass of the $\Dz$ meson is requested to lie within the interval $[1851,1880]\mevcc$, corresponding to about twice the resolution around the known $\Dz$ mass~\cite{PDG2020}.   A large fraction of secondary $D^{\ast+}$ mesons is removed by demanding that the measured $\mathrm{IP}$ of $\Dz$ mesons does not exceed $50\mum$ (see Sect.~\ref{sect:backgrounds}).  This requirement is also very effective at removing combinatorial background. The resolution on the $\Dz$ decay time is improved by performing a kinematic fit~\cite{Hulsbergen:2005pu} in which the $D^{\ast+}$ candidate is required to originate from the PV. The reconstructed $\Dz$ decay time is selected in the interval $[1.0,8.0]\tau_{\Dz}$. The lower bound is chosen to minimise biasing effects related to the differences of the time resolution between the three $\Dz$ decay channels, and to avoid significant combinatorial background from the PV. The higher bound is set to minimise the contribution from secondary decays, the fraction of which increases as a function of $\Dz$ decay time. 

\begin{figure}[tb]
    \centering
    \includegraphics[width=\linewidth]{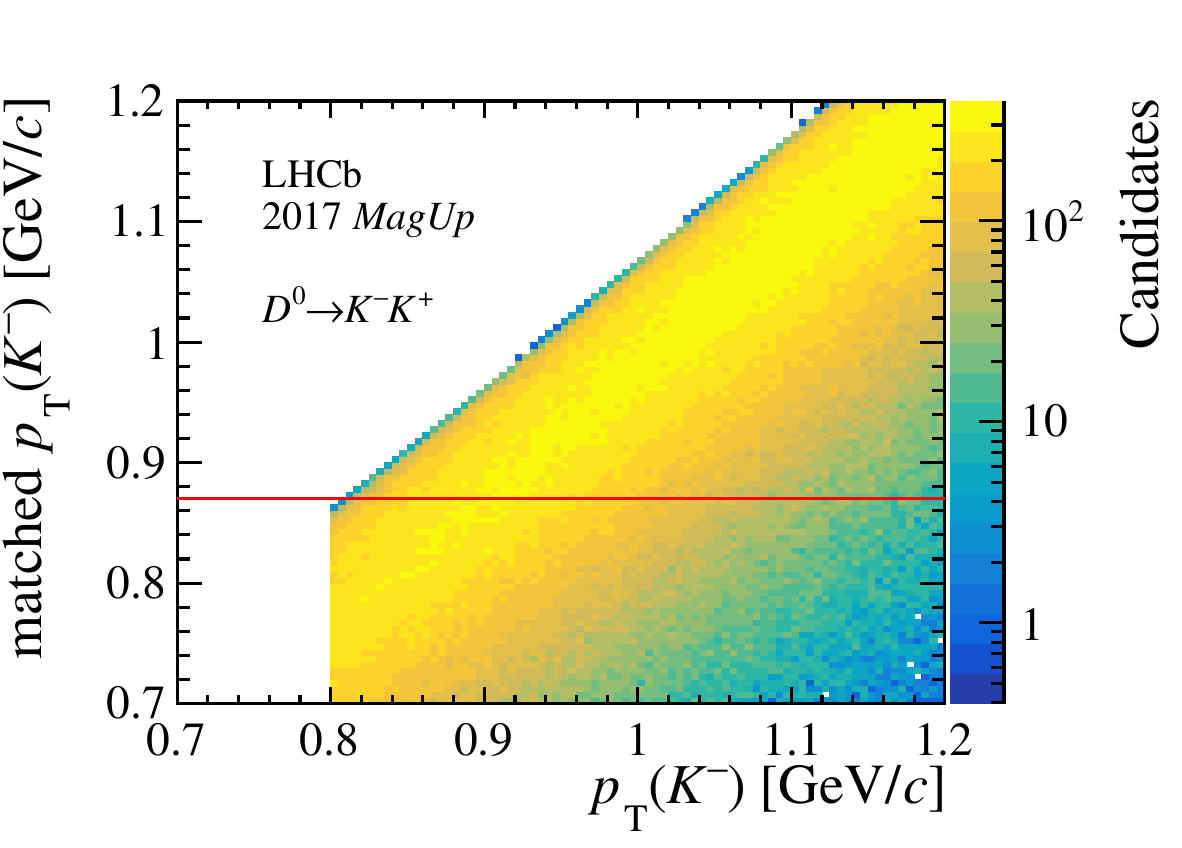}
    \caption{Matched versus original transverse momenta for the matching of a $K^-$ to a $\pi^-$ particle, related to the $\ycp^{CC}$ measurement. The red line represents the requirement applied to the data sample, where candidates below the line are rejected. The plot is obtained with the 2017 \textit{MagUp} sample.}
    \label{fig:h2_p1_PT_p1_PT_matched_LOGZ_KK2PP}
\end{figure}

The kinematic matching procedure is then performed for the selected candidates, as detailed in Sect.~\ref{sect:analysis_strategy}. Figure~\ref{fig:h2_p1_PT_p1_PT_matched_LOGZ_KK2PP} illustrates the transverse momentum of the $K^-$ candidate of a \DzKK decay matched to the $\pi^-$ candidate of a \DzPP decay.
The trigger selection requirement on the kaon transverse momentum at $0.8\gevc$ is visible as a sharp cut on the $x$~axis.
A requirement on the matched transverse momentum of the kaon, visible on the $y$~axis, to be larger than $0.87\gevc$, is effectively tighter than the trigger requirement applied on the \DzKK candidates. The application of this tighter requirement in the selection of both the matched \DzKK and the \DzPP candidates ensures that both decays are selected with the same efficiency profile.
Similarly, for each kinematic variable of the $\Dz$ candidates,
a tightened requirement on the matched variable is applied to the matched and target decays.
For the three measurements described in this paper, the matched (target) decay is that with the smallest (largest)
momentum of the final-state particles in the \Dz rest frame,
which consists in matching kaon to pion candidates, allowing for the minimal loss of statistical precision. Hence, for the $\ycp^{CC}$ measurement, the \DzKK decay is matched to the \DzPP decay; for the $\ycp^{KK} - \ycp^{K\pi}$ measurement, the \DzKK decay is matched to the \DzRS decay; finally, for the $\ycp^{\pi\pi} - \ycp^{K\pi}$ measurement, the \DzRS decay is matched to the \DzPP decay. 
An additional requirement on matched quantities is that the variable $\tilde{\chi}^2_{\mathrm{IP}}=\mathrm{IP}^2/(11.6 + 23.4/\pt)^2$, where $\pt$ is expressed in $\mathrm{GeV}/c$ and $\mathrm{IP}$ in $\mathrm{\upmu m}$, to be larger than $6.0$~\cite{LHCb-DP-2014-001}. This allows the combinatorial background in the data sample to be reduced further. The data sample is split into 22 intervals of $\Dz$ decay time of equal population, with the exception of the four intervals with the largest decay times containing half of the population of the others.

Following the offline and matching requirements, about $6\%$ of the \DzKK and \DzPP and $3.5 \%$ of the \DzRS candidates are combined with multiple $\pi^+_{\mathrm{tag}}$ candidates to form $D^{\ast +}$ meson candidates. When multiple candidates are present in the event, one is selected randomly.

\section{Mass fit and dominant background contributions}
\label{sect:backgrounds}

\begin{figure}
\centering
\begin{minipage}[b]{0.6\textwidth}
\includegraphics[width=1.\textwidth]{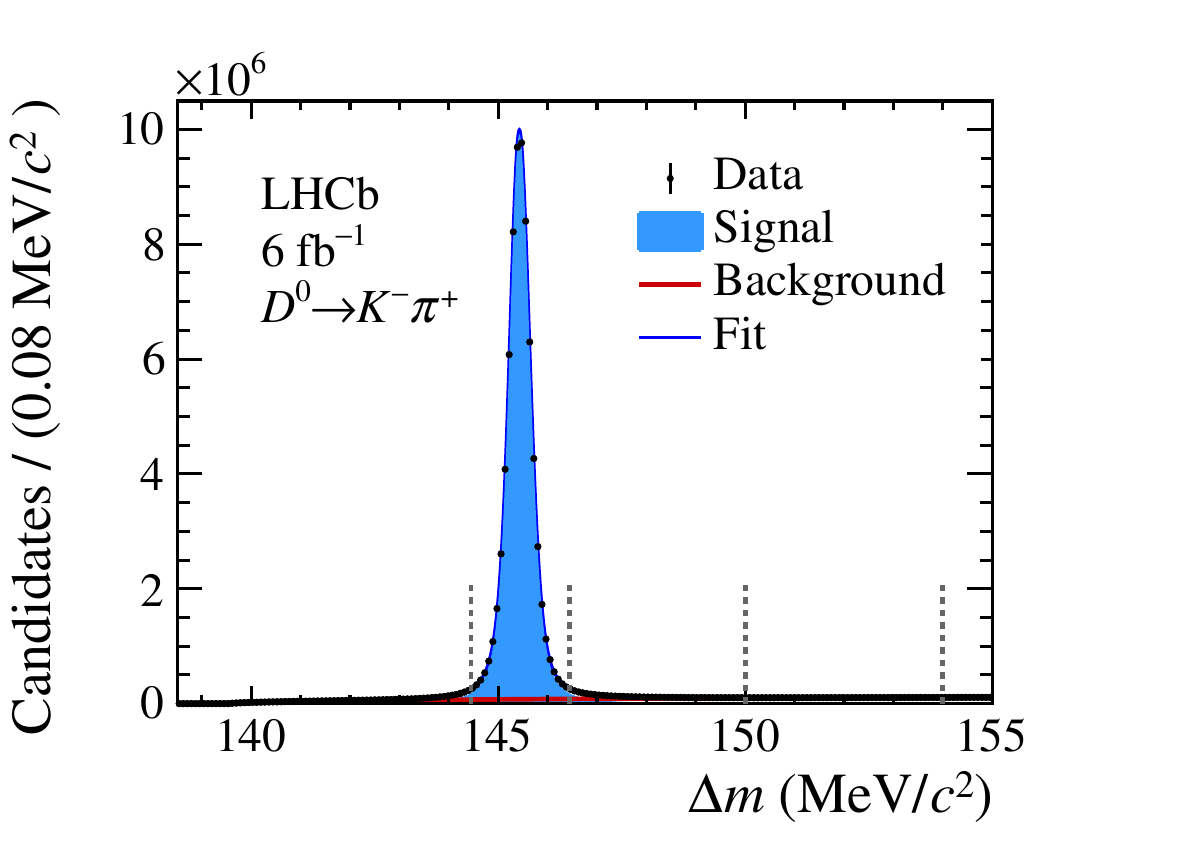}
\end{minipage}
\begin{minipage}[b]{0.6\textwidth}
\includegraphics[width=1.\textwidth]{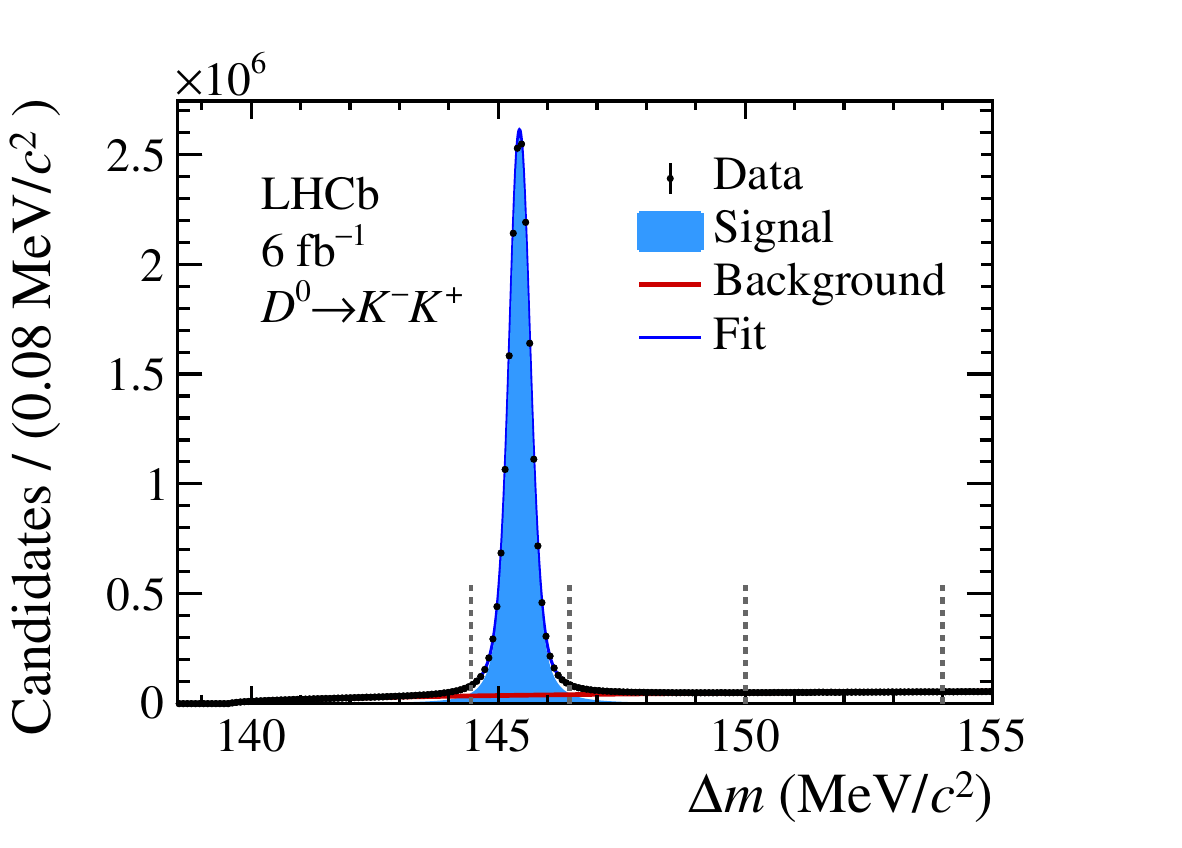}
\end{minipage}
\begin{minipage}[b]{0.6\textwidth}
\includegraphics[width=1.\textwidth]{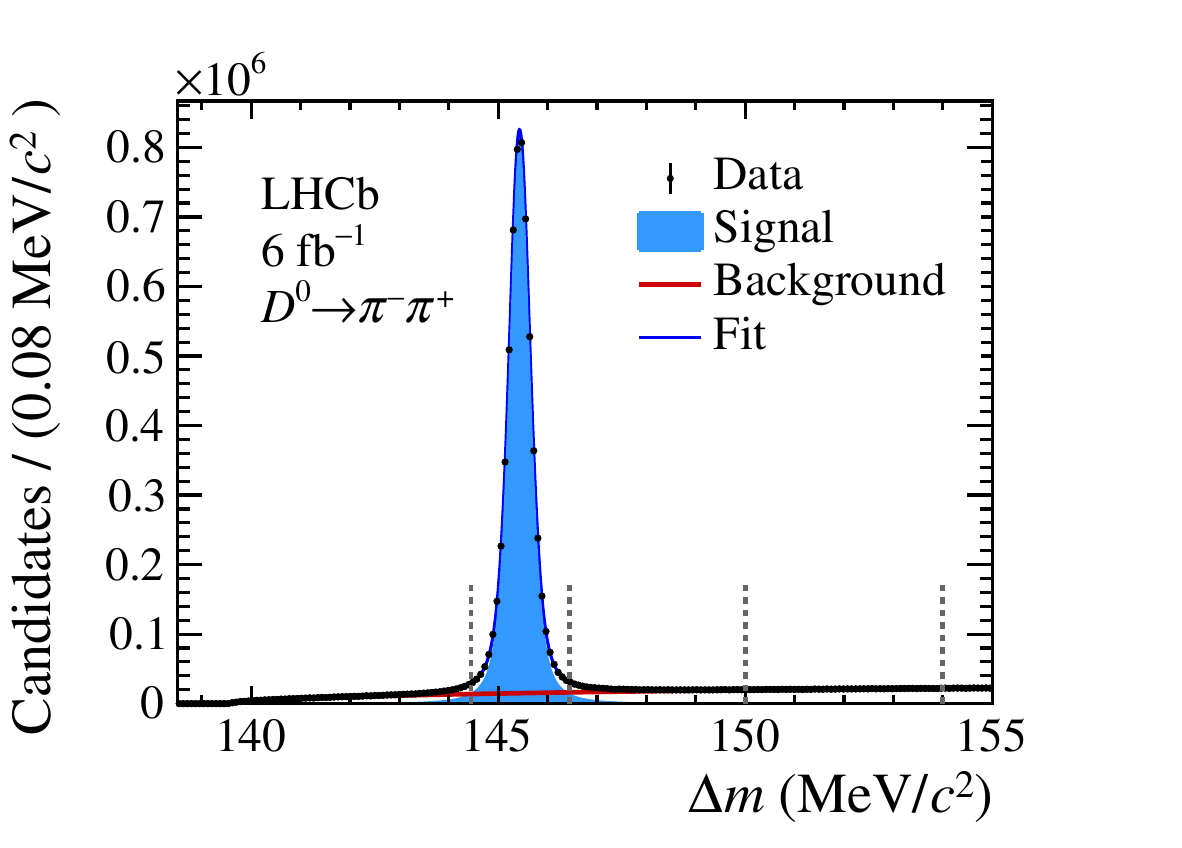}
\end{minipage}
\caption{Distributions of $\Delta m$ for the (top) \DzRS, (centre) \DzKK, and (bottom) \DzPP decay channels for the combined data sample. The signal and sideband regions employed to subtract the combinatorial background are delimited by the dashed vertical lines.
The sum of the fit projections are overlaid.}
\label{fig:dm}
\end{figure}

The $\Delta m$ distributions of all three decay channels are shown in Fig.~\ref{fig:dm} for the combined data set.
A binned maximum-likelihood fit is applied to the $\Delta m$ distribution to separate signal from combinatorial background arising predominantly from the association of a $\Dz$ meson with a random $\pi^{+}_{\mathrm{tag}}$ candidate from the $pp$ interaction.
The signal is fitted with the sum of three Gaussian functions and a Johnson SU function~\cite{Johnson:1949zj}.
The combinatorial background is fitted with the empirical model 
\begin{equation}
\mathcal{P}_{\mathrm{BKG}}(\Delta m |   m_0, \alpha  ) = \frac{1}{\mathcal{I}_B} \Delta m \cdot \sqrt{\frac{\Delta m^2}{m_0^2}-1}\cdot \exp\left(-\alpha \left(\frac{\Delta m^2}{m_0^2}-1\right)\right)\,,
\end{equation}
where $m_0$ and $\alpha$ are free parameters, and $\mathcal{I}_B$ is a normalisation constant.
In the $\Delta m$ distribution, a signal region is defined in the interval $[144.45,146.45]\mevcc$ and a sideband region in the interval $[150,154]\mevcc$. The contribution from the residual background in the signal region is estimated from the sideband region and subtracted with a dedicated procedure. The fitting of the $\Delta m$ distribution is performed independently for each $\Dz$ flavour, year and magnet polarity, and in each of the 22 intervals of $\Dz$ decay time. In the signal region, the time-integrated signal purities are equal to $98\%$, $96\%$, and $95\%$ for the \DzRS, \DzKK and \DzPP channels, respectively, and the time-integrated signal yields amount to $70$ million, $18$ million, and $6$ million decays. The fits to the $\Delta m$ distributions of all three decay channels are displayed in Fig.~\ref{fig:dm}.

\begin{figure}[tb]
\centering
\begin{minipage}[b]{0.495\textwidth}
\includegraphics[width=1.\textwidth]{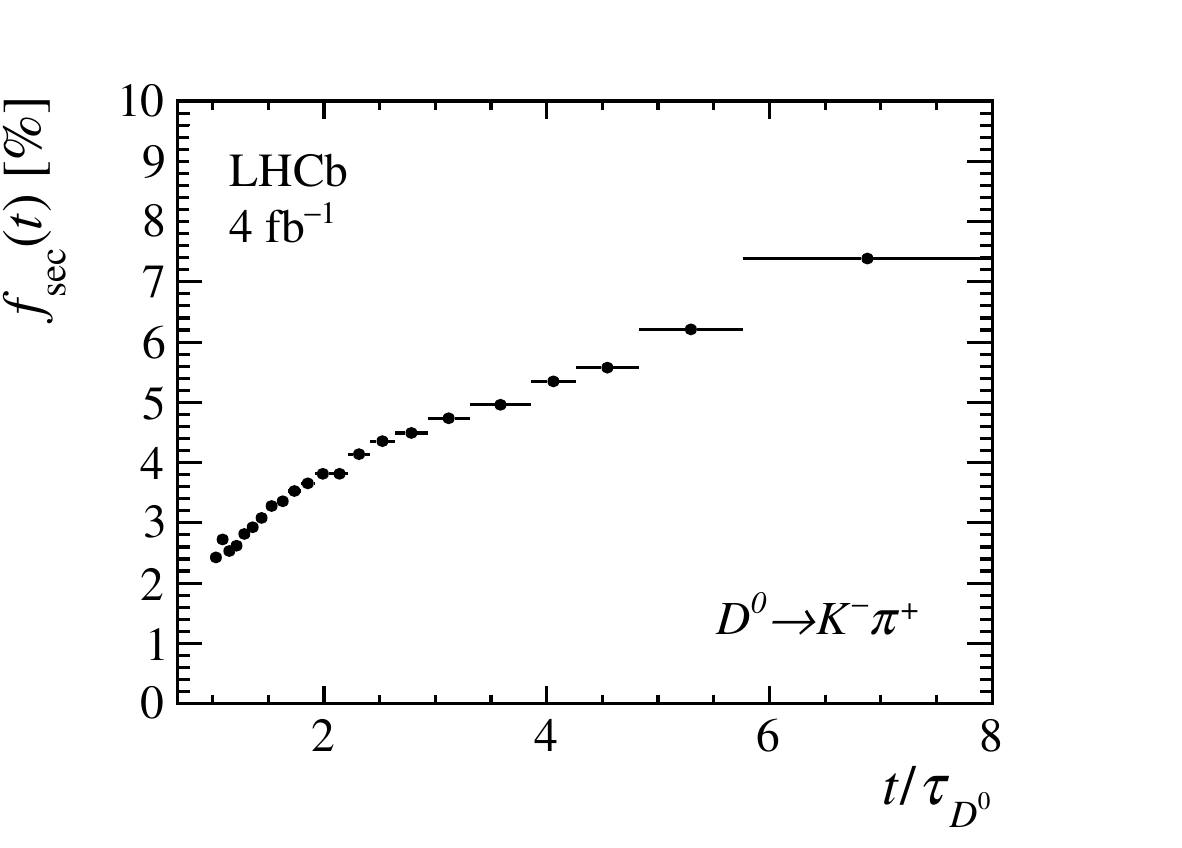}
\end{minipage}
\begin{minipage}[b]{0.495\textwidth}
\includegraphics[width=1.\textwidth]{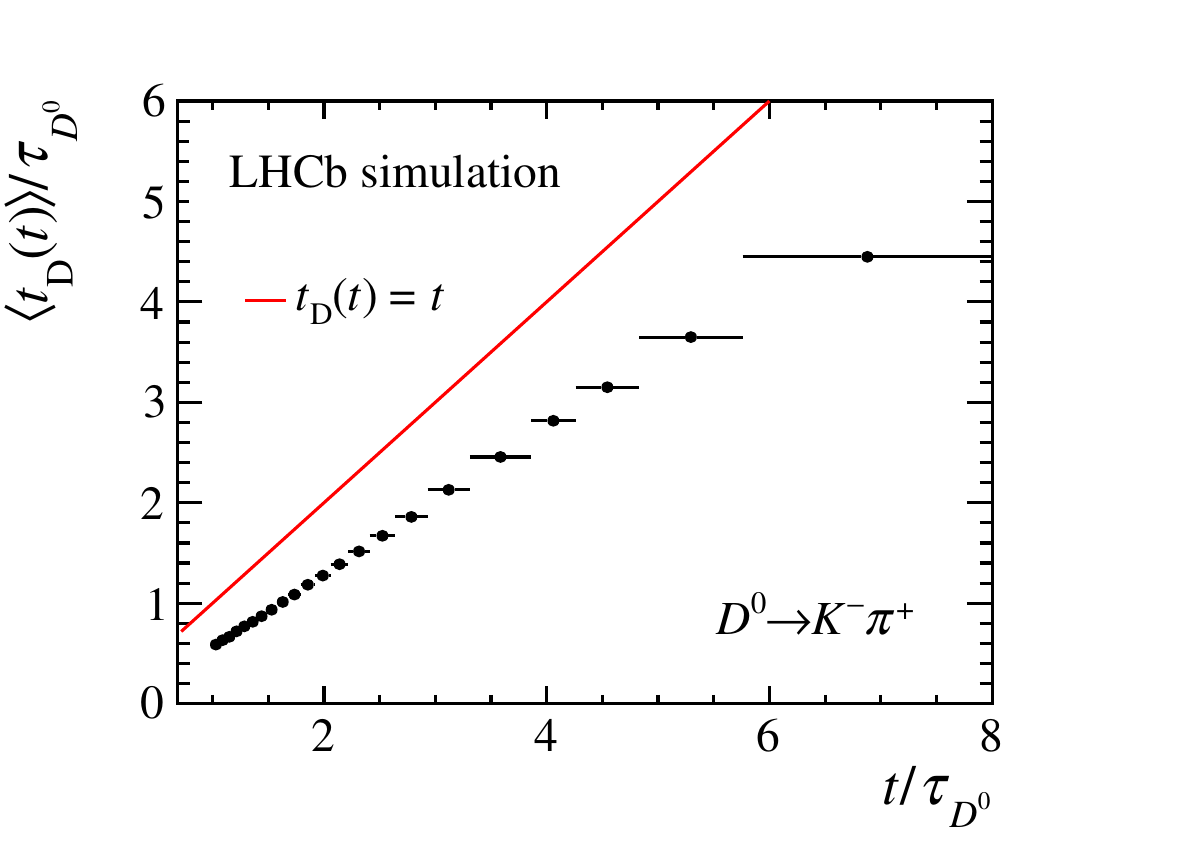}
\end{minipage}
\caption{(Left) fraction and (right) average true $\Dz$ decay time of secondary decays  as a function of the reconstructed $\Dz$ decay time, in units of the average $\Dz$ lifetime.}
\label{fig:fsec_tD}
\end{figure}

The data samples are also contaminated by the presence of secondary $D^{\ast +}$ mesons, which are not produced at the PV but from $B$ meson decays. Since the reconstructed $\Dz$ decay time is calculated as $t=l\cdot m(\Dz)/p(\Dz)$, where $l$ is the measured distance between the PV and the decay vertex of the $\Dz$ meson, $t$ is overestimated for secondary candidates since $l$ is affected by the flight distance of $B$ mesons. The $\mathrm{IP}$ of the corresponding $\Dz$ candidates is usually different from zero, as opposed to $\Dz$ candidates from prompt $D^{\ast +}$ decays. Hence, requesting the $\mathrm{IP}$ of $\Dz$ candidates not to exceed $50\mum$ allows a significant fraction of secondary $D^{\ast +}$ mesons, $f_{\mathrm{sec}}(t) $, defined as the time-dependent ratio of the number of \Dz mesons from secondary decays over the total,
to be rejected from the data sample. To account for the residual contamination of secondary $D^{\ast +}$ candidates, the ratio $R^f(t)$ is separated according to its prompt  and secondary components, $R_{\mathrm{prompt}}^f(t)$ and $R_{\mathrm{sec}}^f(t)$, as
\begin{equation}
    R^f(t) = (1 - f_{\mathrm{sec}}(t))R_{\mathrm{prompt}}^f(t) + f_{\mathrm{sec}}(t)R_{\mathrm{sec}}^f(t) \, .
\label{eq:R_updated_fit_model}
\end{equation}
The decay time ratio of $\Dz$ mesons from secondary $D^{\ast+}$ decays is expressed as
\begin{equation}
    R_{\mathrm{sec}}^f(t) \propto e^{-(\ycp^f - \ycp^{K\pi})\langle t_D(t) \rangle/\tau_{\Dz}} \, , 
\label{eq:Rsec}
\end{equation}
where $\langle t_D(t) \rangle$ is the average true $\Dz$ decay time $\langle t_D \rangle$ as a function of the reconstructed $\Dz$ decay time $t$. The quantities $f_{\mathrm{sec}}(t)$ and $\langle t_D(t) \rangle$ are determined using data and simulated samples of \DzRS decays generated separately for prompt $D^{\ast +}$ decays and through the expected mixture of $B^0$ and $B^{+}$ meson decays to $D^{\ast +}$ candidates. The kinematic distributions of the simulation samples are weighted to those of data samples to account for kinematic discrepancies. 
The fraction $f_{\mathrm{sec}}(t)$ is obtained by fitting the distribution of $\mathrm{IP}(\Dz)$ in data in each interval of $t$ using simulation-based templates of $\mathrm{IP}(\Dz)$ from prompt and secondary decays. The values of $f_{\mathrm{sec}}(t)$ are measured to increase from about $2\%$ to $7\%$ across the studied $\Dz$ decay time range. The quantity $\langle t_D(t) \rangle$ is determined from the simulated sample of secondary decays. The obtained values of $f_{\mathrm{sec}}(t)$ and $\langle t_D(t) \rangle$ are shown in Fig.~\ref{fig:fsec_tD}.

\section{Analysis validation with simulation}
\label{sect:validation_simulation}

The kinematic matching procedure is validated with the use of \rapidsim simulation. Signal candidates of prompt $D^{\ast +} \ra (\DzKK) \pi^+_{\mathrm{tag}}$ and $D^{\ast +} \ra (\DzRS) \pi^+_{\mathrm{tag}}$ decays are generated 
without \Dz--\Dzb mixing.
The simulation samples are subjected to selection criteria representative of the trigger. These include requirements on momentum and $\mathrm{IP}$-related quantities, which are strongly correlated with the $\Dz$ decay time and induce substantial differences between the selection efficiency profiles of \DzKK and \DzRS decays at low $\Dz$ decay time. The kinematic matching procedure is then applied to equalise the selection efficiencies of \DzKK and \DzRS decays.
Following this correction, a fit to the decay time ratio $R^{KK}(t)$ gives $\ycp^{KK} - \ycp^{K\pi} = (0.17 \pm 0.19) \times 10^{-3}$, compatible with the expected value of zero. This study demonstrates that the kinematic matching procedure corrects effectively for the kinematic differences between the two decays.

The analysis procedure is further validated with full simulation. Large signal yields of 50 million \DzRS, 33 million \DzKK and 11 million \DzPP decays are obtained by generating the particles of the studied decay chain without the full underlying event. The analysis procedure detailed in Sect.~\ref{sect:selection} is applied to all three decay channels independently for each year and magnet polarity to account for potential differences between the data taking conditions, and the results are combined as a final step. 
Following the application of the kinematic matching and weighting procedures, the parameters are measured to be 
\begin{equation*}
\begin{split}
    \ycp^{CC} & = (0.15 \pm 0.36) \times 10^{-3} \, , \\
    \ycp^{\pi\pi} - \ycp^{K\pi} & = (0.17 \pm 0.43) \times 10^{-3} \, , \\
    \ycp^{KK} - \ycp^{K\pi} & = (0.10 \pm 0.24) \times 10^{-3} \, ,
\end{split}
\end{equation*}
where the uncertainties are smaller than the statistical uncertainties expected in data.
All three results are compatible with zero. This is expected since \Dz--\Dzb mixing has not been simulated.
This result validates the analysis procedure with simulation. 

\section{Results}

\begin{figure}[tb]
\centering
\begin{minipage}[b]{0.495\textwidth}
\includegraphics[width=1.\textwidth]{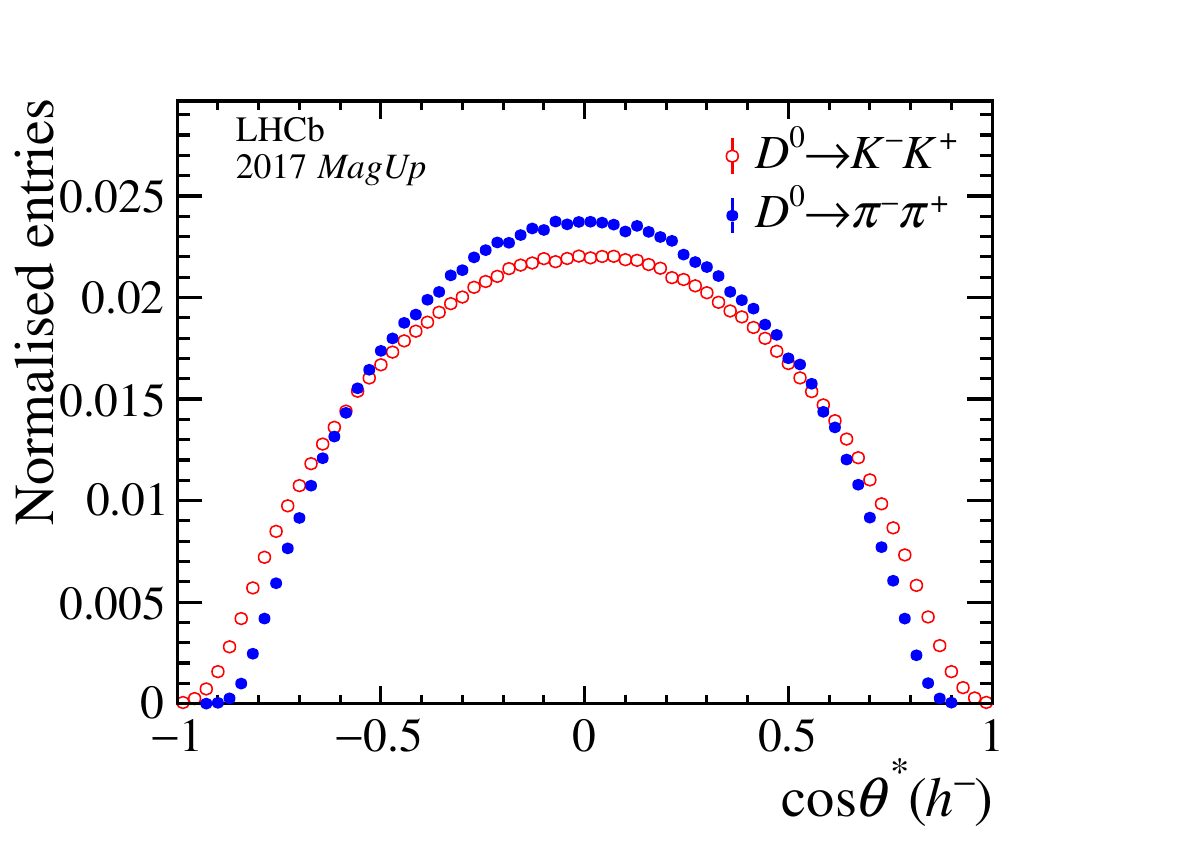}
\end{minipage}
\begin{minipage}[b]{0.495\textwidth}
\includegraphics[width=1.\textwidth]{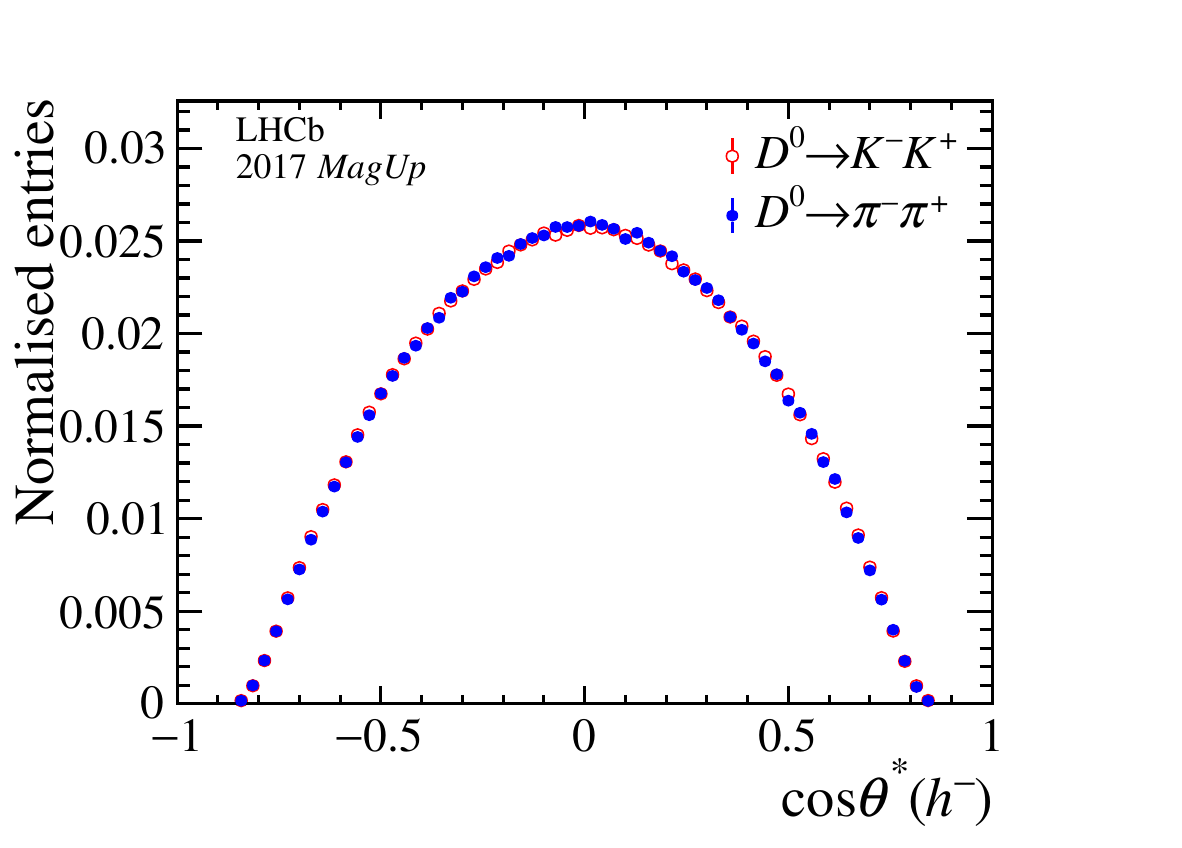}
\end{minipage}
\begin{minipage}[b]{0.495\textwidth}
\includegraphics[width=1.\textwidth]{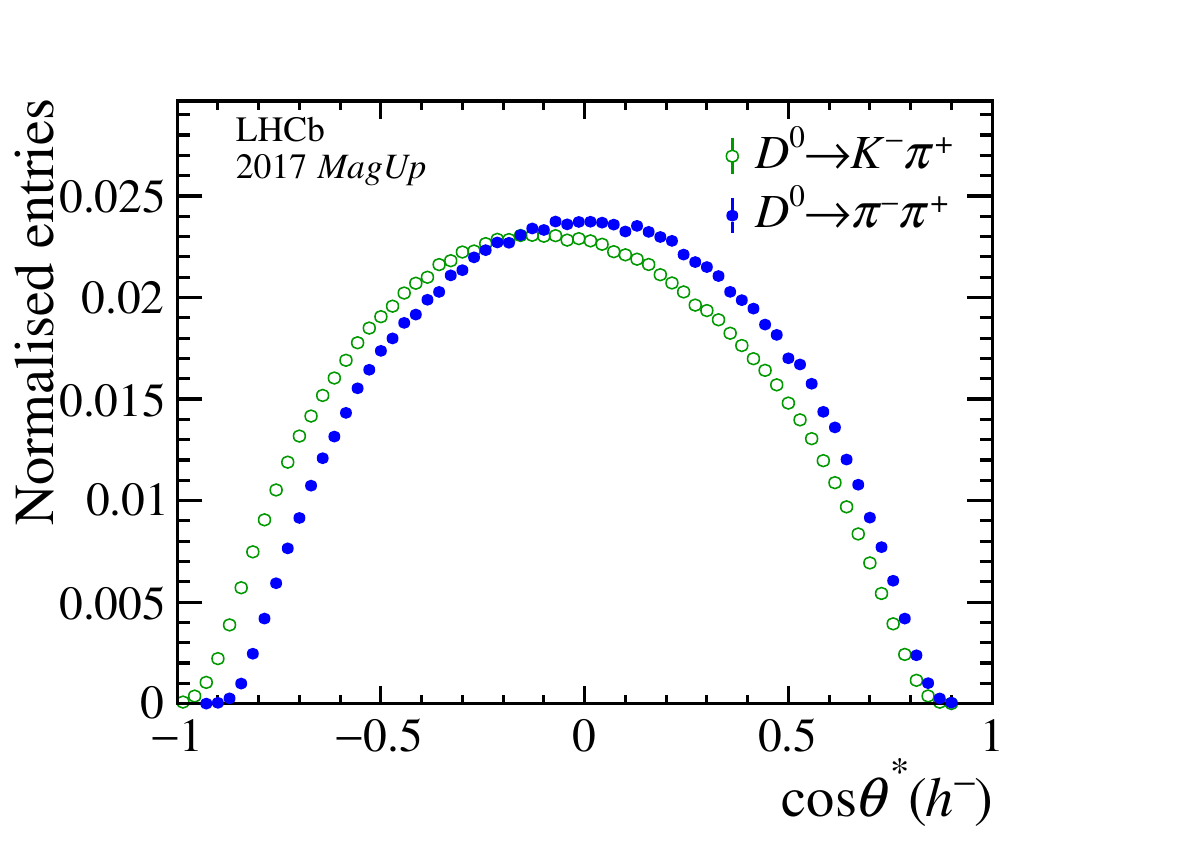}
\end{minipage}
\begin{minipage}[b]{0.495\textwidth}
\includegraphics[width=1.\textwidth]{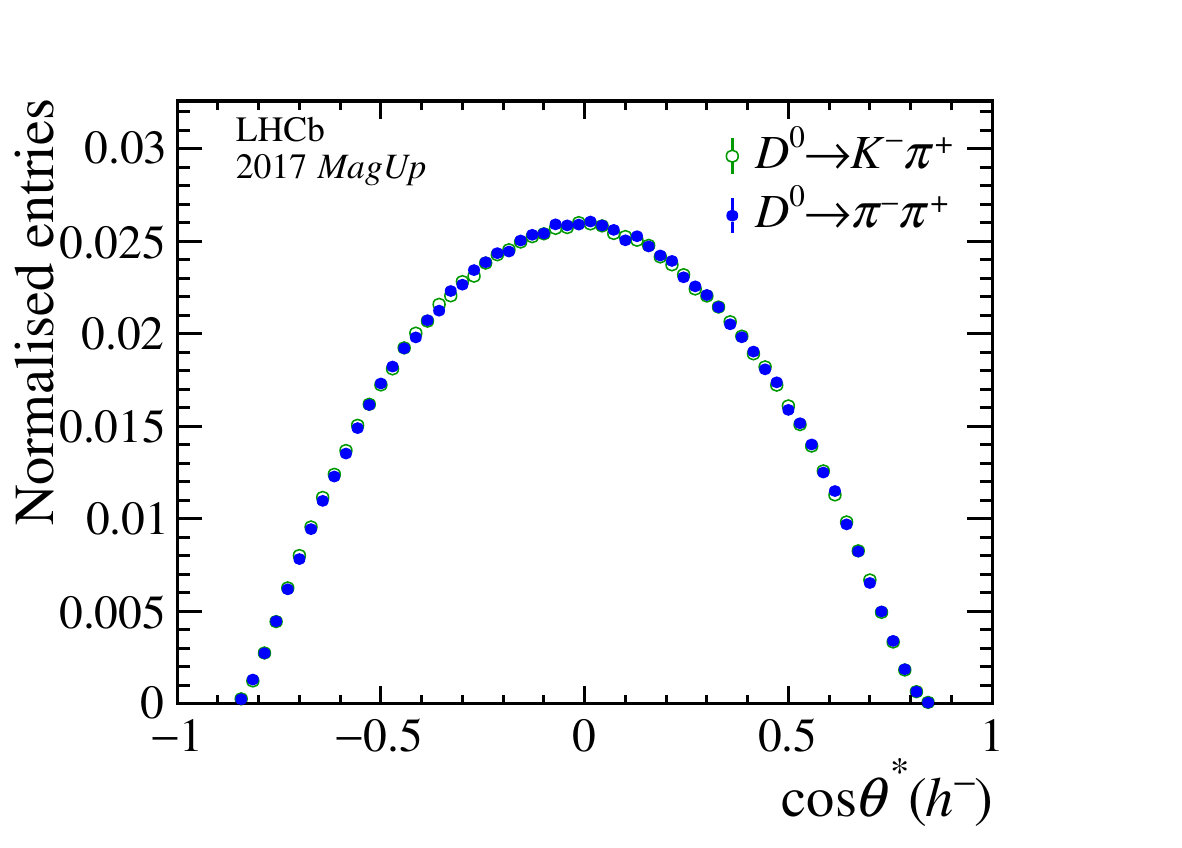}
\end{minipage}
\begin{minipage}[b]{0.495\textwidth}
\includegraphics[width=1.\textwidth]{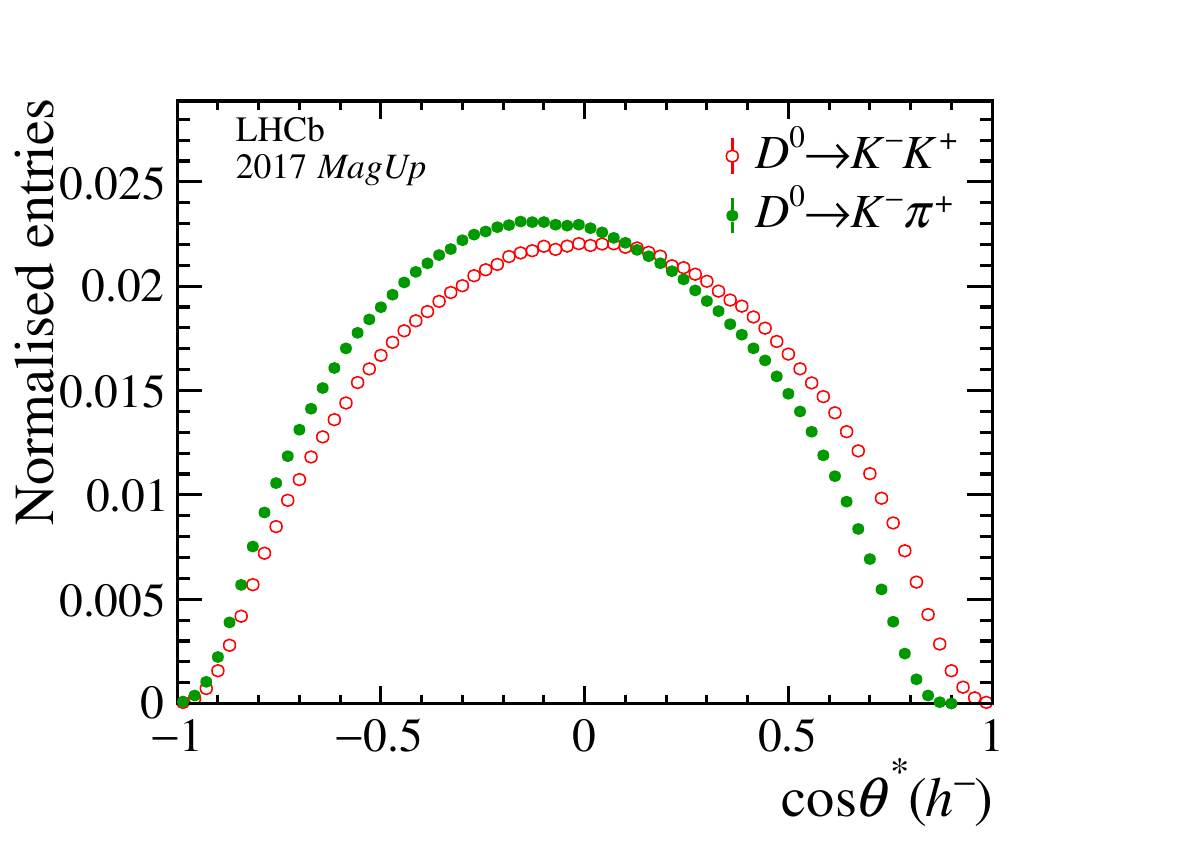}
\end{minipage}
\begin{minipage}[b]{0.495\textwidth}
\includegraphics[width=1.\textwidth]{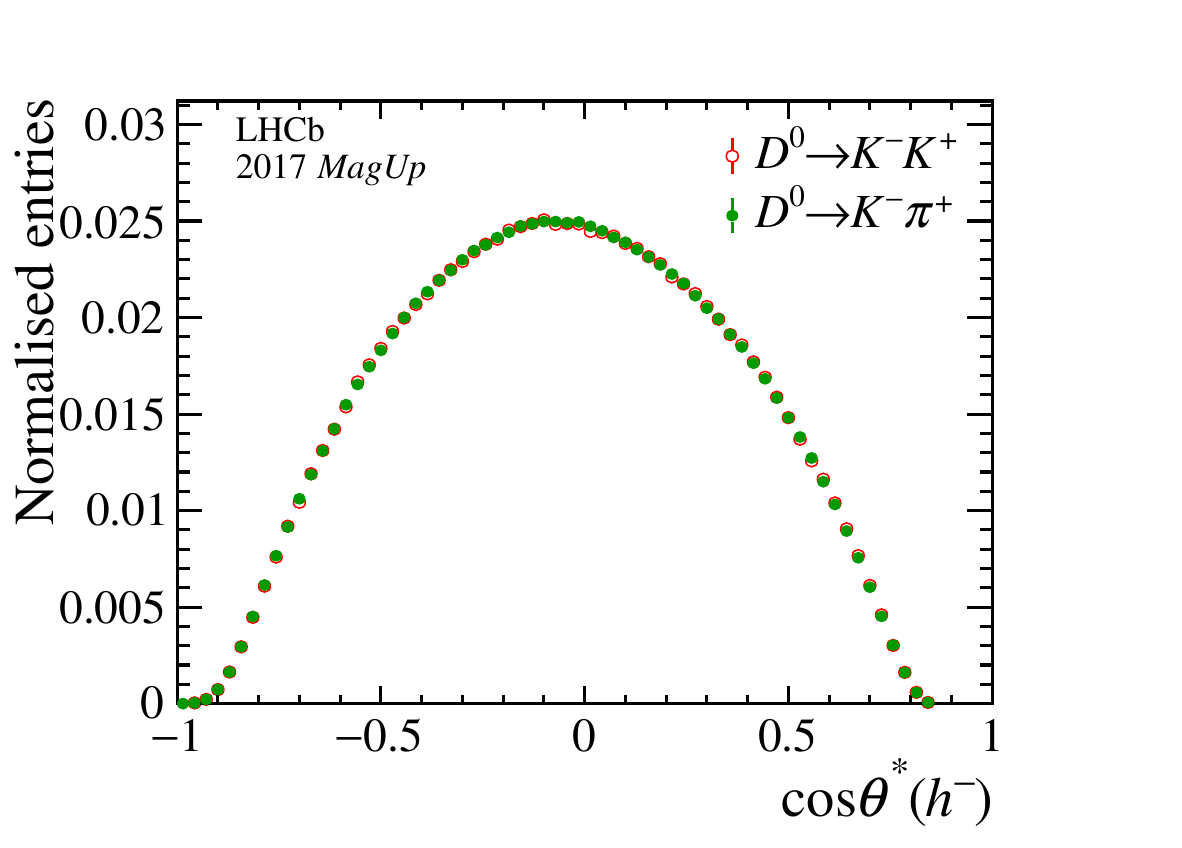}
\end{minipage}
\caption{(Left) normalised distributions of the $D^0$ decay angle $\cos \theta^{\ast}(h^-)$ in the raw condition,  and (right) following both kinematic matching and reweighting procedures. 
The distributions are shown for the  (top) $\ycp^{CC}$, (middle) $\ycp^{\pi\pi} - \ycp^{K\pi}$ and (bottom) $\ycp^{KK} - \ycp^{K\pi}$ measurements. The plots are obtained with the 2017 \textit{MagUp} sample.}
\label{fig:yCP_kinematics_costhetastar}
\end{figure}

Both matching and weighting procedures are employed to perform the measurements of $\ycp^{CC}$, $\ycp^{\pi\pi} - \ycp^{K\pi}$ and $\ycp^{KK} - \ycp^{K\pi}$ for each year and magnet polarity of the \lhcb Run~2 data set. Figure~\ref{fig:yCP_kinematics_costhetastar} presents the normalised distributions of the $D^0$ decay angle prior to any kinematic correction (raw) and after the application of both kinematic matching and weighting procedures. The two correction procedures significantly improve the agreement between the distributions. The agreement is also verified to be good for a series of additional kinematic variables.

\begin{figure}
\centering
\begin{minipage}[b]{0.6\textwidth}
\includegraphics[width=1.\textwidth]{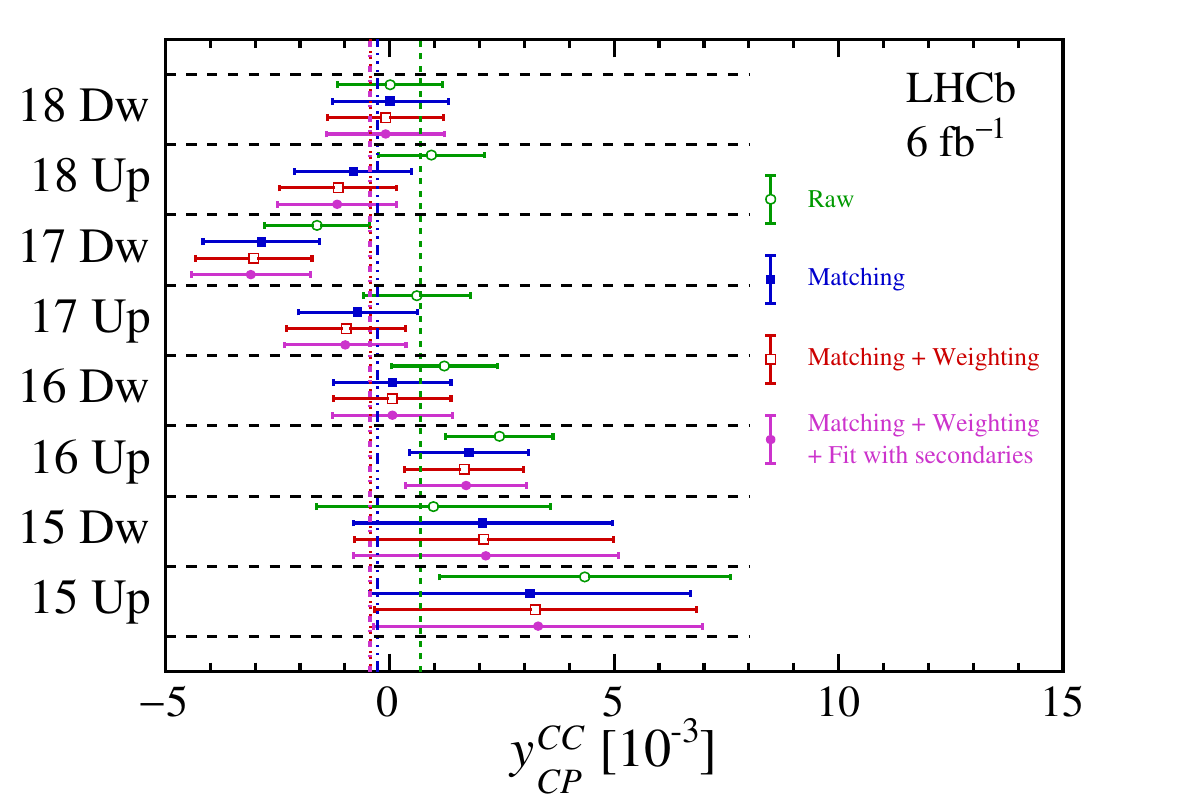}
\end{minipage}
\begin{minipage}[b]{0.6\textwidth}
\includegraphics[width=1.\textwidth]{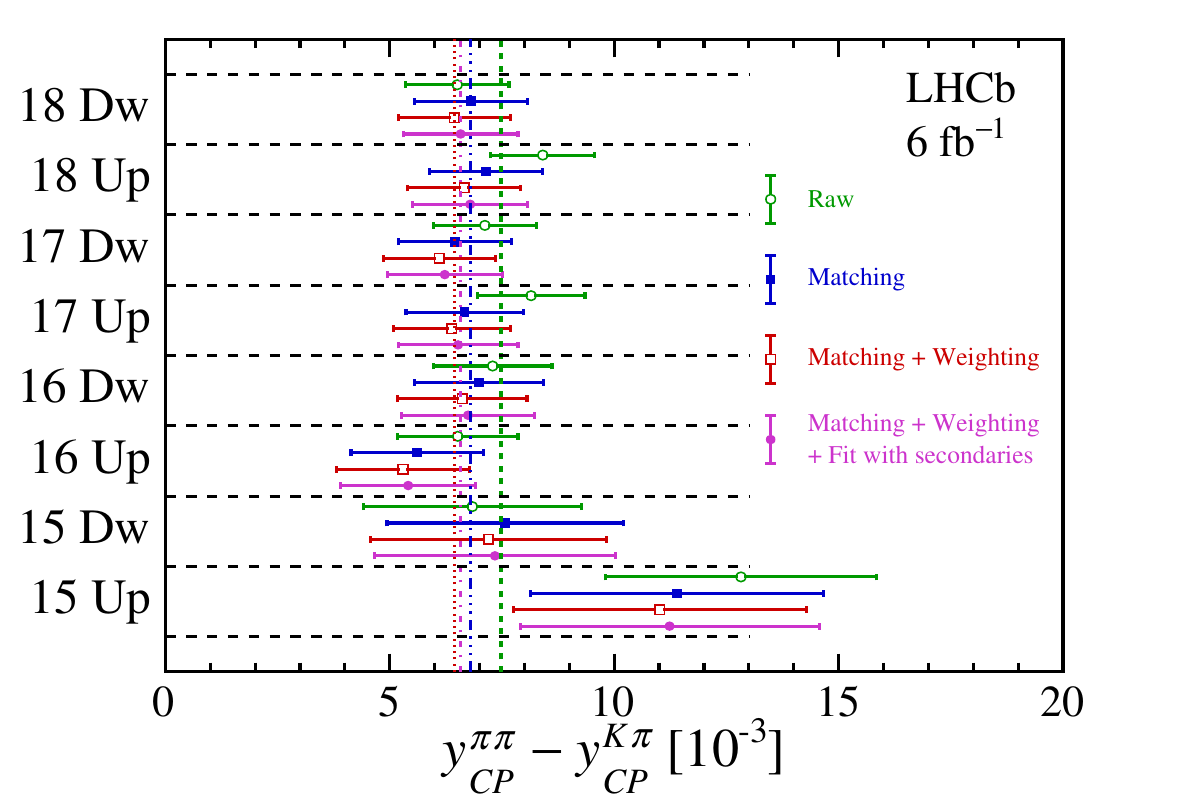}
\end{minipage}
\begin{minipage}[b]{0.6\textwidth}
\includegraphics[width=1.\textwidth]{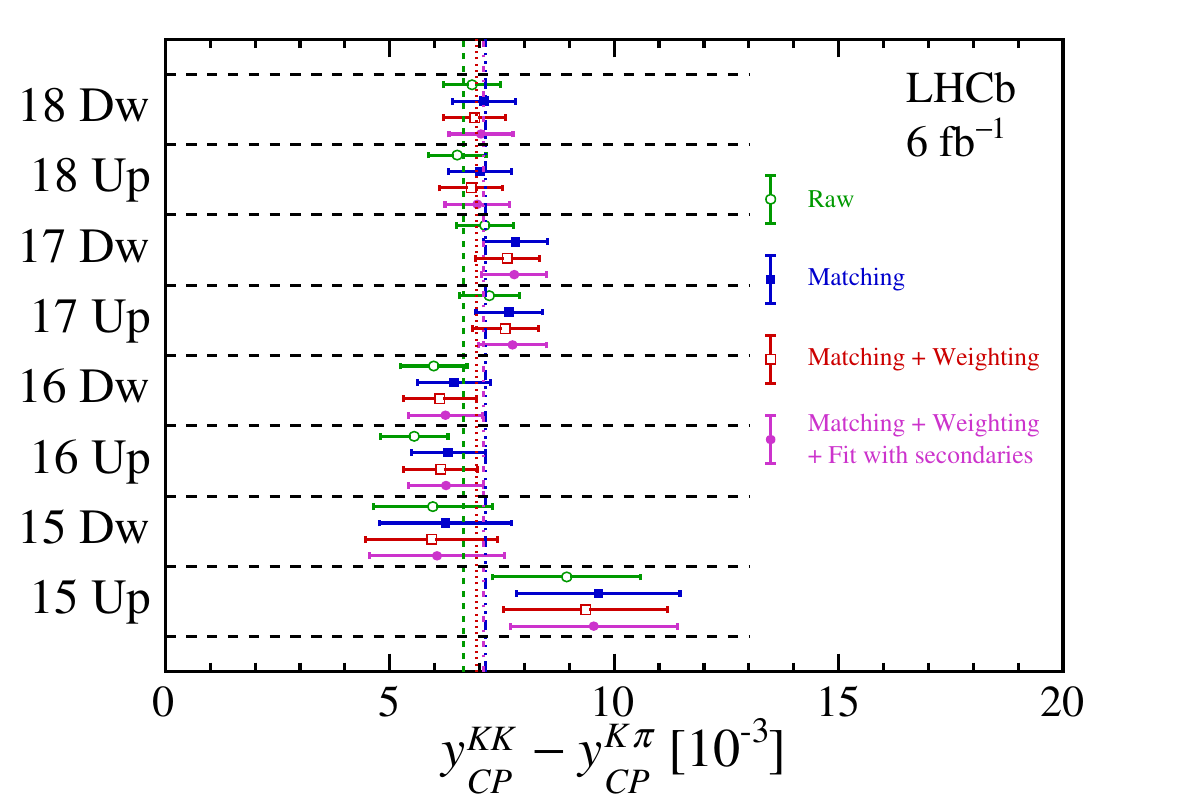}
\end{minipage}
\caption{Results for (top) $\ycp^{CC}$, (centre) $\ycp^{\pi\pi} - \ycp^{K\pi}$ and (bottom) $\ycp^{KK} - \ycp^{K\pi}$. The measurements employing raw data, and following both matching and weighting conditions are shown in green, blue and red, respectively. The measurements in purple employ the fit model where the presence of secondary decays is considered. The dashed vertical lines correspond to $\chi^2$ fits, used to determine the average values other all subsamples.
In the y-axis labels, the data-taking year is abbreviated with the last two digits only and the magnet polarity \MagUp (\MagDown) is abbreviated as ``Up'' (``Dw'').}
\label{fig:yCP_Full_Run2}
\end{figure}

The parameters $\ycp^{CC}$, $\ycp^{KK} - \ycp^{K\pi}$, and $\ycp^{\pi\pi} - \ycp^{K\pi}$ are determined from a $\chi^2$ fit to the corresponding time-dependent $R^f(t)$ ratios.
The results of the measurements
are presented in Fig.~\ref{fig:yCP_Full_Run2}, where $\chi^2$ fits with a constant function are performed to determine the averages over all data samples. The results of these fits are reported in Table~\ref{tab:yCP_results}.
The raw measurements have good compatibility among the different years and magnet polarities. This indicates uniform performance of the trigger and offline selections, which do not include effects substantially biasing the measurements. The kinematic matching procedure shifts the average value of $\ycp^{CC}$ by $(-0.96 \pm 0.21) \times 10^{-3}$, $\ycp^{\pi\pi} - \ycp^{K\pi}$ by $(-0.67 \pm 0.21) \times 10^{-3}$, and $\ycp^{KK} - \ycp^{K\pi}$ by $(+0.50 \pm 0.12) \times 10^{-3}$. The shifts of $\ycp^{\pi\pi} - \ycp^{K\pi}$ and $\ycp^{KK} - \ycp^{K\pi}$ are compatible in magnitude but opposite in sign, as expected given the difference in the nature of the final states in the numerators of their respective decay time ratios. The shifts of $\ycp^{CC}$ are measured to be about twice those of $\ycp^{\pi\pi} - \ycp^{K\pi}$, expected from the fact that $R^{CC}(t)$ probes the decay time ratio of final states in which both particles have different masses. 
The kinematic weighting shifts the values of $\ycp^{CC}$ and $\ycp^f - \ycp^{K\pi}$ by a few $10^{-4}$ towards negative values. 
Finally, the use of the fit model of Eq.~\eqref{eq:R_updated_fit_model}, which takes into account the presence of secondary decays, shifts the average values of $\ycp^{f} - \ycp^{K\pi}$ by about $+0.1\times 10^{-3}$. 

\begin{figure}[tb]
\centering
\begin{minipage}[b]{0.495\textwidth}
\includegraphics[width=1.\textwidth]{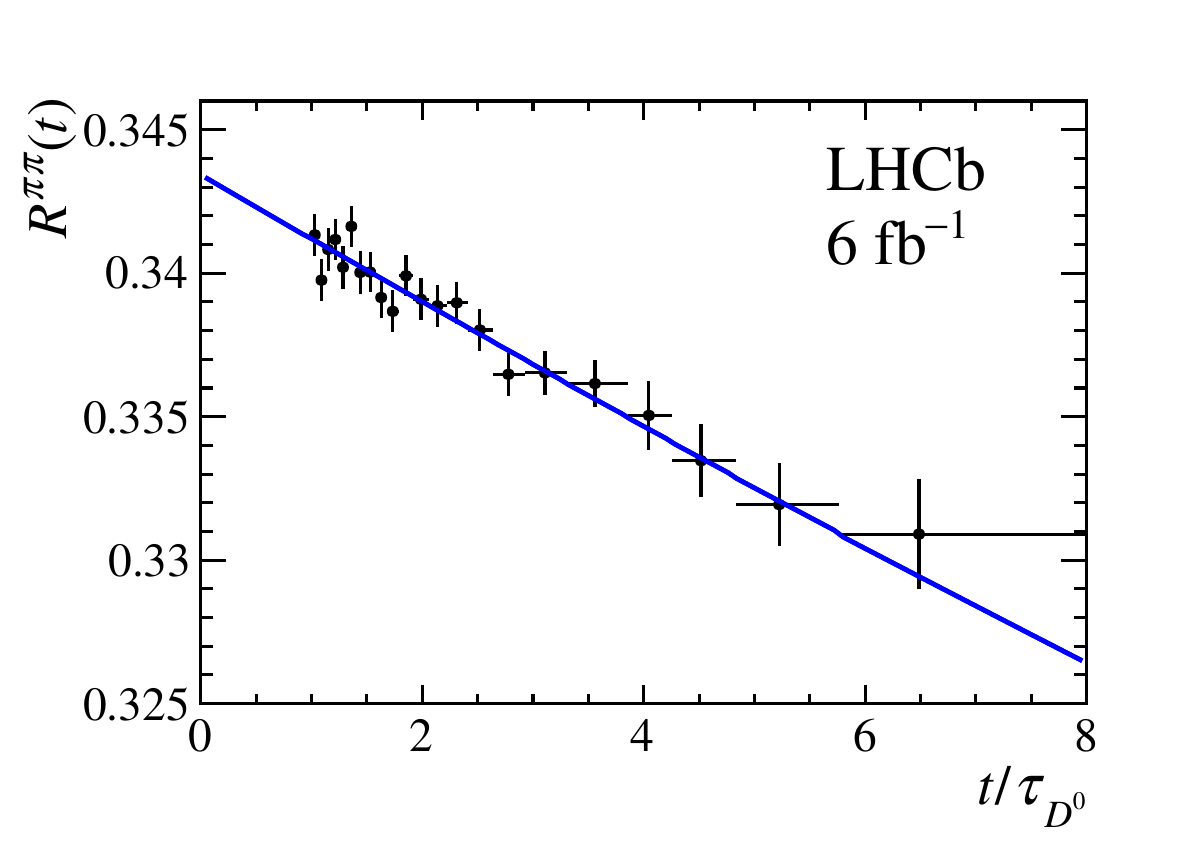}
\end{minipage}
\begin{minipage}[b]{0.495\textwidth}
\includegraphics[width=1.\textwidth]{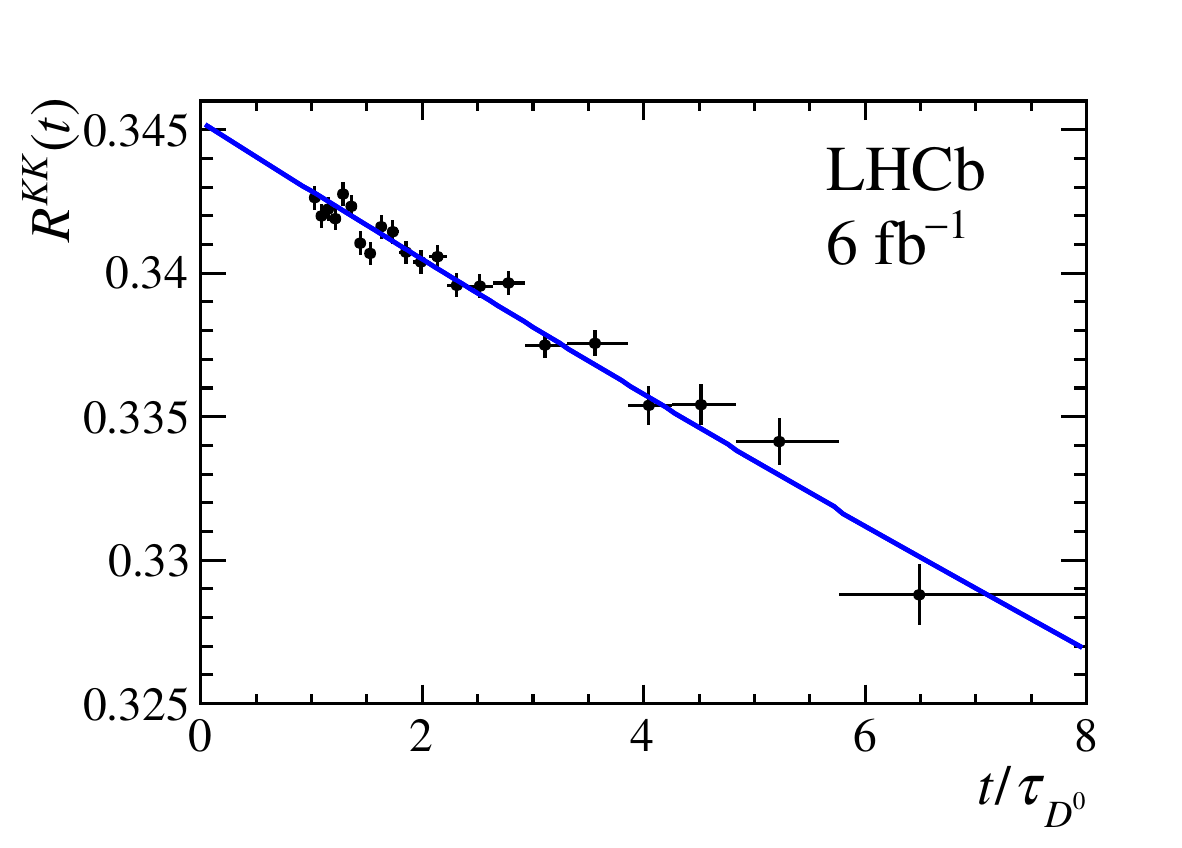}
\end{minipage}
\caption{Distributions of  (left) $R^{\pi\pi}(t)$ and (right)  $R^{KK}(t)$ using the full \lhcb Run~2 data set, with the results of $\ycp^{\pi\pi} - \ycp^{K\pi}$ and $\ycp^{KK} - \ycp^{K\pi}$ overlaid as the blue slopes.}
\label{fig:yCP_allYears_Run2}
\end{figure}

All three measurements have individual fits of good quality and are found to be compatible among years and magnet polarities. Following all correction procedures and the use of the fit model of Eq.~\eqref{eq:R_updated_fit_model}, which includes secondary decays, the average values are measured to be
\begin{equation*}
\begin{split}
    \ycp^{CC} & = (-0.44 \pm 0.53) \times 10^{-3} \, , \\
    \ycp^{\pi\pi} - \ycp^{K\pi} & = (6.57 \pm 0.53) \times 10^{-3} \, , \\
    \ycp^{KK} - \ycp^{K\pi} & = (7.08 \pm 0.30) \times 10^{-3} \, ,
\end{split}
\label{eq:yCP_results_Run2}
\end{equation*}
where the uncertainties are only statistical.
The value of $\ycp^{CC}$ is measured to be compatible with zero within one standard deviation ($\sigma$), validating the cross-check measurement with data. 
Figure~\ref{fig:yCP_allYears_Run2} shows the distributions of $R^{\pi\pi}(t)$ and $R^{KK}(t)$ using the full data set, with Eq.~\eqref{eq:R_updated_fit_model} overlaid, computed using the average values of $\ycp^{\pi\pi} - \ycp^{K\pi}$ and $\ycp^{KK} - \ycp^{K\pi}$. 

\begin{table}[tb]
    \centering
    \def\arraystretch{1.2}
    \begin{tabular}{cccc}
    \hline
     & $\ycp^{CC}$ & $\ycp^{KK}-\ycp^{K\pi}$ & $\ycp^{KK}-\ycp^{K\pi}$ \\
    \hline
    \multirow{2}{*}{Raw} & \multirow{2}{*}{$0.68 \pm 0.47 \: (7.9)$} & \multirow{2}{*}{$7.48 \pm 0.48 \: (5.5)$} & \multirow{2}{*}{$6.64 \pm 0.27 \: (6.6)$} \\ \\
    \hline
    \multirow{2}{*}{Matching} & \multirow{2}{*}{$-0.28 \pm 0.52 \: (8.3)$} & \multirow{2}{*}{$6.80 \pm 0.52 \: (2.9)$} & \multirow{2}{*}{$7.14 \pm 0.29 \: (5.5)$} \\ \\
    \hline
    \multirow{2}{*}{Matching + Weighting} & \multirow{2}{*}{$-0.43 \pm 0.52 \: (9.0)$} & \multirow{2}{*}{$6.44 \pm 0.52 \: (2.8)$} & \multirow{2}{*}{$6.94 \pm 0.29 \: (5.9)$} \\  \\
    \hline
    Matching + Weighting & \multirow{2}{*}{$-0.44 \pm 0.53 \: (9.0)$} & \multirow{2}{*}{$6.57 \pm 0.53 \: (2.8)$} & \multirow{2}{*}{$7.08 \pm 0.30 \: (5.9)$} \\
    + Fit with secondaries & & & \\
    \hline
    \end{tabular}
    \caption{Results of the $\chi^2$ fits of Fig.~\ref{fig:yCP_Full_Run2} for each correction procedure. The results are shown in units of $10^{-3}$, while the values in parenthesis correspond to the $\chi^2$ of the fits, where the number of degrees of freedom is 7 for all measurements.}
    \label{tab:yCP_results}
\end{table}

\section{Systematic uncertainties}
\label{sect:systematics}

Although the analysis procedure is designed to minimise systematic uncertainties on $\ycp^{f} - \ycp^{K\pi}$, several sources of possible bias in the results remain and are evaluated.
The first source of systematic uncertainty arises from the subtraction of the combinatorial background, which relies on the assumption that the kinematic properties of combinatorial background candidates are identical in the signal and in the sideband interval of the $\Delta m$ distribution. A systematic uncertainty on this
assumption is assigned by repeating the measurement using three alternative sideband
regions, namely $[140.5,142]$, $[149,152] $ and $[152,155]\mevcc$. An additional source of systematic uncertainty is assigned by propagating the uncertainties on the combinatorial background subtraction. The combined systematic uncertainty is measured to be  $0.12 \times 10^{-3}$ for $\ycp^{\pi\pi} - \ycp^{K\pi}$  and $0.07\times 10^{-3}$ for $\ycp^{KK} - \ycp^{K\pi}$, the first being  higher due to larger combinatorial background level and smaller size of the \DzPP sample.

A second source of systematic uncertainty is related to the presence of partially reconstructed or misreconstructed $D^{\ast +} \ra \Dz\pi^+$ decays. These decays are referred to as peaking background since they show a peaking structure in the $\Delta m$ distribution. 
For each decay channel, the peaking background contributions are studied with \rapidsim simulation. The simulation samples are used as templates to fit the $\Dz$ mass distributions, from which the fraction of peaking background candidates is determined in the signal region.
In the \DzPP channel, the  $\Dz \ra \pi^-e^+\nu_{e}$ and $\Dz \ra \pi^-{\mu}^+\nu_{\mu}$ background components are significant in the $\Dz$ signal mass region. The background fraction is measured as $3.5 \times 10^{-4}$. 
In the \DzRS channel, the $\Dz \ra \pi^-\pi^+\pi^0$, $\Dz \ra K^-e^+\nu_e$ and $\Dz \ra K^-\mu^+\nu_{\mu}$ background contributions are dominant, and the background fraction is estimated as $8.9 \times 10^{-4}$. Finally, in the \DzKK channel, the dominant background sources come from the $\Dz \ra K^-\pi^+\pi^0$, $\Dz \ra K^-e^+\nu_e$ and $\Dz \ra K^-\mu^+\nu_{\mu}$ decay channels. The background fraction is measured to be $11.8 \times 10^{-4}$. 
Using the \rapidsim samples, the impact of these contributions on the measurements of $\ycp^{\pi\pi} - \ycp^{K\pi}$ and $\ycp^{KK} - \ycp^{K\pi}$ is evaluated to be $0.02 \times 10^{-3}$ and $0.11 \times 10^{-3}$, respectively. These values are assigned as systematic uncertainties.

A third source of systematic uncertainty arises from the precision on the determination of the time-dependent fraction of secondary decays $f_{\mathrm{sec}}(t)$, and the average true $\Dz$ decay time $\langle t_D \rangle$ as a function of the reconstructed $\Dz$ decay time $t$.
The uncertainty in the determination of $f_{\mathrm{sec}}(t)$
receives three separate contributions. First, in the fits to $\mathrm{IP}(\Dz)$, discrepancies in the ratio between the fit model and data are seen to reach up to $10\%$. The impact of these discrepancies to the measurement is assigned as a systematic uncertainty.
Second, the fits to $\mathrm{IP}(\Dz)$ are performed in the interval $[0,200]\mum$.
The impact of increasing the upper bound of the interval to $600\mum$ results in a small variation to the $\ycp^{f} - \ycp^{K\pi}$ measurements, which is assigned as a systematic uncertainty.
Finally, the simulation samples of prompt and secondary candidates are produced for the 2017 and 2018 data conditions only, and a systematic uncertainty is assigned by considering the impact to the measurement of potential variations to the distribution of $\mathrm{IP}(\Dz)$ in the 2015 and 2016 data taking conditions. 
The uncertainty in the determination of $\langle t_D(t) \rangle$ receives two independent contributions. First, the impact on the difference of lifetimes between $B^0$ and $B^+$ mesons is considered by determining $\langle t_D \rangle (t)$ using simulation samples of alternatively only $B^0$ or $B^+$ candidates. Then, the effect of the weighting of the simulation samples is evaluated by determining $\langle t_D(t) \rangle$ with and without the weighting. 
The total systematic uncertainty related to the treatment of secondary decays is evaluated to be at the level of $0.03\times 10^{-3}$ for both the $\ycp^{\pi\pi} - \ycp^{K\pi}$ and the $\ycp^{KK} - \ycp^{K\pi}$ measurements. 

A systematic uncertainty related to the kinematic weighting procedure is assigned by using alternative input variables and particles of the decay chain to the weighting algorithm and repeating the measurement. 
An additional systematic uncertainty is assigned by performing the weighting of the target decay to the matched one. The systematic uncertainties are summed in quadrature and determined as $0.08 \times 10^{-3}$ for $\ycp^{\pi\pi} - \ycp^{K\pi}$ and $0.02\times 10^{-3}$ for $\ycp^{KK} - \ycp^{K\pi}$.

The uncertainty on the current world average value of the lifetime of the $\Dz$ meson~\cite{PDG2020} is propagated as a systematic uncertainty and is estimated as $0.03 \times 10^{-3}$ for both measurements.

A source of systematic uncertainty includes a potential bias related to the contributions from the flavour of the $\Dz$ meson, which can arise from tagging-pion detection asymmetries and $D^{\ast+}$ production asymmetries. The size of such a bias is estimated by performing the measurement separating the two $\Dz$ flavours. The $\Dz$ and $\Dbar^0$ measurements are seen to be compatible within the $1\sigma$ level. The weighted average of the $\Dz$ and $\Dbar^0$ measurements is compared to the baseline measurement where both $\Dz$ and $\Dbar^0$ samples are merged. For $\ycp^{\pi\pi} - \ycp^{K\pi}$, the difference between the two strategies is measured as $0.03\times 10^{-3}$, while for $\ycp^{KK} - \ycp^{K\pi}$ it is found to be below $0.01\times 10^{-3}$. These values are assigned as systematic uncertainty.

It is suggested in Ref.~\cite{Pajero_Morello_paper} that in the expansion of $R^f(t)$, the second-order terms of the decay time and mixing parameters  differ from those of the exponential function used in the baseline fit. To estimate potential resulting biases, one thousand pseudoexperiments consisting of samples of Cabibbo-suppressed and Cabibbo-favoured decays are generated according to the theoretical model described in Ref.~\cite{Pajero_Morello_paper}, using the current world average values of charm mixing and \CP-violation parameters~\cite{LHCb-PAPER-2021-033}. For each pseudoexperiment, $R^f(t)$ is fitted with an exponential function and the departure of the mean value of the fits from the expectation of the theoretical model is measured as $0.03 \times 10^{-3}$, and is assigned as a systematic uncertainty. 
The summary of the systematic uncertainties is presented in Table~\ref{tab:systematics_summary}, where the quadratic sum of all contributions is $0.16 \times 10^{-3}$ for $\ycp^{\pi\pi} - \ycp^{K\pi}$ and $0.14 \times 10^{-3}$ for $\ycp^{KK} - \ycp^{K\pi}$. 
\begin{table}[tb]
    \centering
    \def\arraystretch{1.2}
    \begin{tabular}{lcc}
    \hline
     & $\sigma(\ycp^{\pi\pi} - \ycp^{K\pi})$ & $\sigma(\ycp^{KK} - \ycp^{K\pi})$ \\
     & $[10^{-3}]$ & $[10^{-3}]$ \\
    \hline
    Combinatorial background & $0.12$ & $0.07$ \\
    Peaking background & $0.02$ & $0.11$ \\
    Treatment of secondary decays & $0.03$ & $0.03$ \\
    Kinematic weighting procedure & $0.08$ & $0.02$ \\
    Input $\Dz$ lifetime & $0.03$ & $0.03$ \\
    Residual nuisance asymmetries & $0.03$ & $<0.01$ \\ 
    Fit bias & $0.03$ & $0.03$ \\
    \hline
    Total & $0.16$ & $0.14$ \\
    \hline
    \end{tabular}
    \caption{Systematic uncertainties for the $\ycp^{\pi\pi} - \ycp^{K\pi}$ and $\ycp^{KK} - \ycp^{K\pi}$ measurements.}
    \label{tab:systematics_summary}
\end{table}

Robustness checks are performed by verifying that the measurements do not show any dependence on various variables, including the momentum, transverse momentum, the pseudorapidity and azimuthal angle of the $\Dz$ and $\pi_{\mathrm{tag}}^+$ mesons, as well as the $\Dz$ flight distance in the plane transverse to the beam and the $z$ coordinate of the $\Dz$ decay vertex. No significant dependence of $\ycp^f - \ycp^{K\pi}$ on any of the listed variables is observed. To study a potential dependence on the orientation of the magnetic field and on a potential left-right asymmetry of the detector, the measurement is performed separately for positive and negative values of the $x$ component of the momentum of the $\Dz$ meson, and for the \textit{MagUp} and \textit{MagDown} polarities. All measurements are seen to be statistically compatible within two standard deviations. To evaluate the impact of possible residual resolution effects at low values of $\Dz$ decay time, the measurements of $\ycp^f - \ycp^{K\pi}$ are performed by increasing  successively the lower window of the $\Dz$ decay time up to $1.7 \, \tau_{\Dz}$. Accounting for the statistical overlap, all measured values are found to be statistically compatible and correspondingly no systematic uncertainty is assigned to this effect.

Due to the presence of correlations between the reconstructed $\Dz$ decay time and momentum, the correction procedure can introduce a bias to the true values of $\ycp^f - \ycp^{K\pi}$. To study this bias, artificial values of $\ycp^f - \ycp^{K\pi}$ are injected to the data samples by altering the decay time distribution of numerator decays. Both kinematic matching and weighting procedures are applied and the measured values are compared to the injected ones. The procedure is performed for ten values in the interval  $[-25,25]\times 10^{-3}$. The measured and injected values agree, confirming that no significant bias is seen and correspondingly no systematic uncertainty is applied.

\section{Summary and conclusion}
\label{sect:results}

The measurements of the ratios of the effective decay widths of \mbox{\DzPP} and \mbox{\DzKK} decays over that of \DzRS decays are performed with the \lhcb experiment using \proton\proton collisions at a centre-of-mass energy of $13\tev$, corresponding to an integrated luminosity of $6\invfb$.
The ratios give direct access to the charm mixing parameters $\ycp^{\pi\pi} - \ycp^{K\pi}$ and $\ycp^{KK} - \ycp^{K\pi}$, which are measured to be 
\begin{equation*}
\begin{split}
    \ycp^{\pi\pi} - \ycp^{K\pi} & = (6.57 \pm 0.53 \pm 0.16) \times 10^{-3} \, , \\
    \ycp^{KK} - \ycp^{K\pi} & = (7.08 \pm 0.30 \pm 0.14) \times 10^{-3} \, ,
\end{split}
\label{eq:yCP_results_Run2_with_sys}
\end{equation*}
where the first uncertainties are statistical and the second systematic. Assuming that all systematic uncertainties are fully correlated, except those of the peaking background contributions which are considered as uncorrelated, the combination of the two measurements yields
\begin{equation*}
    \ycp - \ycp^{K\pi} = (6.96 \pm 0.26 \pm 0.13) \times 10^{-3} \, .
\end{equation*}
This result is compatible with the present world average~\cite{HFLAV18} and more precise by a factor of four.

A combination of  \lhcb charm measurements is performed using the statistical framework detailed in Ref.~\cite{LHCb-PAPER-2021-033}.
When the present result is added, the mixing parameter $y$ is found to be equal to $y=\left(6.46 \pm ^{+0.24}_{-0.25}\right)\times 10^{-3}$, improving its current sensitivity by more than a factor of two~\cite{HFLAV18}. In addition, the strong phase difference between the $D^0 \to K^{\mp}\pi^{\pm}$ decay amplitudes is $\delta_{K\pi} = \left(192.1^{+3.7}_{-4.0}\right)^{\circ}$ and departs from $180^{\circ}$ by about three standard deviations, indicating an evidence for $U$-spin symmetry breaking. 

The precision on $y$ and $\delta_{K\pi}$ can be further reduced by a simultaneous combination of charm results with measurements of the Cabibbo-Kobayashi-Maskawa angle $\gamma$ in beauty decays, as first done in Ref.~\cite{LHCb-PAPER-2021-033}. This will be the subject of a separate publication.

\section*{Acknowledgements}
%
%
\noindent We express our gratitude to our colleagues in the CERN
accelerator departments for the excellent performance of the LHC. We
thank the technical and administrative staff at the LHCb
institutes.
We acknowledge support from CERN and from the national agencies:
CAPES, CNPq, FAPERJ and FINEP (Brazil); 
MOST and NSFC (China); 
CNRS/IN2P3 (France); 
BMBF, DFG and MPG (Germany); 
INFN (Italy); 
NWO (Netherlands); 
MNiSW and NCN (Poland); 
MEN/IFA (Romania); 
MSHE (Russia); 
MICINN (Spain); 
SNSF and SER (Switzerland); 
NASU (Ukraine); 
STFC (United Kingdom); 
DOE NP and NSF (USA).
We acknowledge the computing resources that are provided by CERN, IN2P3
(France), KIT and DESY (Germany), INFN (Italy), SURF (Netherlands),
PIC (Spain), GridPP (United Kingdom), RRCKI and Yandex
LLC (Russia), CSCS (Switzerland), IFIN-HH (Romania), CBPF (Brazil),
PL-GRID (Poland) and NERSC (USA).
We are indebted to the communities behind the multiple open-source
software packages on which we depend.
Individual groups or members have received support from
ARC and ARDC (Australia);
AvH Foundation (Germany);
EPLANET, Marie Sk\l{}odowska-Curie Actions and ERC (European Union);
A*MIDEX, ANR, IPhU and Labex P2IO, and R\'{e}gion Auvergne-Rh\^{o}ne-Alpes (France);
Key Research Program of Frontier Sciences of CAS, CAS PIFI, CAS CCEPP, 
Fundamental Research Funds for the Central Universities, 
and Sci. \& Tech. Program of Guangzhou (China);
RFBR, RSF and Yandex LLC (Russia);
GVA, XuntaGal and GENCAT (Spain);
the Leverhulme Trust, the Royal Society
 and UKRI (United Kingdom).



\newpage
\addcontentsline{toc}{section}{References}
\bibliographystyle{LHCb}
\bibliography{main,standard,LHCb-PAPER,LHCb-CONF,LHCb-DP,LHCb-TDR}

\newpage
\centerline
{\large\bf LHCb collaboration}
\begin
{flushleft}
\small
R.~Aaij$^{32}$,
A.S.W.~Abdelmotteleb$^{56}$,
C.~Abell{\'a}n~Beteta$^{50}$,
F.~Abudin{\'e}n$^{56}$,
T.~Ackernley$^{60}$,
B.~Adeva$^{46}$,
M.~Adinolfi$^{54}$,
H.~Afsharnia$^{9}$,
C.~Agapopoulou$^{13}$,
C.A.~Aidala$^{87}$
% ...

{\footnotesize \it

% ...
$^{9}$Universit{\'e} Clermont Auvergne, CNRS/IN2P3, LPC, Clermont-Ferrand, France\\
% ...
$^{13}$LPNHE, Sorbonne Universit{\'e}, Paris Diderot Sorbonne Paris Cit{\'e}, CNRS/IN2P3, Paris, France\\
% ...
$^{32}$Nikhef National Institute for Subatomic Physics, Amsterdam, Netherlands\\
% ...
$^{46}$Instituto Galego de F{\'\i}sica de Altas Enerx{\'\i}as (IGFAE), Universidade de Santiago de Compostela, Santiago de Compostela, Spain\\
% ...
$^{50}$Physik-Institut, Universit{\"a}t Z{\"u}rich, Z{\"u}rich, Switzerland\\
% ...
$^{54}$H.H. Wills Physics Laboratory, University of Bristol, Bristol, United Kingdom\\
% ...
$^{56}$Department of Physics, University of Warwick, Coventry, United Kingdom\\
% ...
$^{60}$Oliver Lodge Laboratory, University of Liverpool, Liverpool, United Kingdom\\
% ...
$^{68}$Syracuse University, Syracuse, NY, United States\\
% ...
$^{87}$University of Michigan, Ann Arbor, United States, associated to $^{68}$\\
\bigskip
% ...
}
\end{flushleft}

\end{document}